\documentclass[10pt]{iopart}
\usepackage{graphicx,epsfig}% Include figure files
\usepackage{bm}% bold math
\usepackage{iopams}
\newcommand{\ba}{\begin{eqnarray}}
\newcommand{\ea}{\end{eqnarray}}
\newcommand{\bse}{\numparts}
\newcommand{\ese}{\endnumparts}

\newcommand{\DD}{{\cal {D}}}

\newcommand{\bbq}{\begin{quote}}
\newcommand{\eeq}{\end{quote}}
\newcommand{\tbb}{t_{\textrm{\tiny{bb}}}}

\newcommand{\tls}{t_{\textrm{\tiny{LS}}}}
\newcommand{\Hls}{H_{\textrm{\tiny{LS}}}}

\newcommand{\rhols}{\rho_{\textrm{\tiny{LS}}}}
\newcommand{\CR}{{\cal{R}}}
\newcommand{\RR}{{\cal{R}}^{(3)}}

\newcommand{\T}{{}^3{\cal{T}}}

\newcommand{\EE}{{\cal{E}}}

\newcommand{\HH}{{\cal{H}}}
\newcommand{\HHav}{\langle{\cal{H}}\rangle}
\newcommand{\KK}{{\cal{K}}}

\newcommand{\QQ}{{\cal{Q}}}
\newcommand{\MM}{{\cal{M}}}

\newcommand{\Aav}{\langle A\rangle}

\newcommand{\Omav}{\langle\Omega\rangle}

\newcommand{\PhiNL}{\Phi_{\textrm{\tiny{NL}}}}

\newcommand{\DENL}{{\textrm{\bf{D}}}_{\textrm{\tiny{NL}}}}

\newcommand{\Da}{\delta^{(A)}}
\newcommand{\Daav}{\delta_{\textrm{\tiny{NL}}}^{(A)}}

\newcommand{\Dh}{\delta^{(\HH)}}
\newcommand{\Dhav}{\delta_{\textrm{\tiny{NL}}}^{(\HH)}}

\newcommand{\Dk}{\delta^{(k)}}

\newcommand{\rhoav}{\langle\rho\rangle}
\newcommand{\Drho}{\delta^{(\rho)}}
\newcommand{\Drhoav}{\delta_{\textrm{\tiny{NL}}}^{(\rho)}}

\newcommand{\KKav}{\langle{\cal{K}}\rangle}
\newcommand{\DKK}{\delta^{(\KK)}}
\newcommand{\DKKav}{\delta_{\textrm{\tiny{NL}}}^{(\KK)}}

\newcommand{\DOm}{\delta^{(\Omega)}}

\newcommand{\dd}{{\rm{d}}}
\newcommand{\Del}{{\textrm{\bf{D}}}}

\begin{document}

\title[Weighed scalar averaging in LTB dust models, part II]{Weighed scalar averaging in LTB dust models, part II: a formalism of exact perturbations. } 
\author{ Roberto A. Sussman}
\address{Instituto de Ciencias Nucleares, Universidad Nacional Aut\'onoma de M\'exico (ICN-UNAM),
A. P. 70--543, 04510 M\'exico D. F., M\'exico.}
\eads{$^\ddagger$\mailto{sussman@nucleares.unam.mx}}
\date{\today}
\begin{abstract} We examine the exact perturbations that arise from the q--average formalism that was applied in the preceding article (part I) to Lema\^\i tre--Tolman--Bondi (LTB) models. By introducing an initial value parametrization, we show that all LTB scalars that take a FLRW ``look alike'' form (frequently used in the literature dealing with LTB models) follow as q--averages of covariant scalars that are common to FLRW models. These q--scalars determine for every averaging domain a unique FLRW background state through Darmois matching conditions at the domain boundary, though  the definition of this background does not require an actual matching with a FLRW region (Swiss cheese type models). Local perturbations describe the deviation from the FLRW background state through the local gradients of covariant scalars at the boundary of every comoving domain, while non--local perturbations do so in terms of the intuitive notion of a ``contrast'' of local scalars with respect to FLRW reference values that emerge from q--averages assigned to the whole domain or the whole time slice in the asymptotic limit.  We derive fluid flow evolution equations that completely determine the dynamics of the models in terms of the q--scalars and both types of perturbations. A rigorous formalism of exact spherical non--linear perturbations is defined over the FLRW background state associated to the q--scalars, recovering the standard results of linear perturbation theory in the appropriate limit. We examine the notion of the amplitude and illustrate the differences between local vs non--local perturbations by qualitative diagrams and through an example of a cosmic density void that follows from the numeric solution of the evolution equations.                                             
\end{abstract}

\maketitle
\section{Introduction.}

in the preceding article (part I) we introduced for the study of LTB models \cite{LTB} a formalism based on a new set of scalar variables (the q--scalars) that follow from applying a weighed proper volume average (q--average) to covariant fluid flow LTB scalars that are common with FLRW models. As proven in part I, the q--scalars are not coordinate \"ansatze, but covariant scalars related to curvature and kinematic invariants, and thus provide an elegant and coordinate independent representation of the models that is alternative to the standard variables normally used in the literature \cite{KH1,KH2,KH3,KH4,BoKrHe,ltbstuff,kras3,focus} (see reviews in \cite{focus,kras1,kras2,BKHC2009}). All proper curvature and kinematic tensors characteristic of the models are expressible in terms of irreducible algebraic expansions formed with the metric and 4--velocity, whose coefficients are local fluctuations of these scalars. Also, all scalar invariant contractions of these tensors are quadratic fluctuations of the q--scalars whose q--averages are statistical moments (variance and covariance) of the density and Hubble scalar expansion.   

As shown in part I (see summary in section 2), the q--scalars can be, either functionals defined on arbitrary fixed domains, or functions (``q--functions'') when considering the pointwise dependence of the average on the varying boundary of a domain. By comparing q--scalars with the non--averaged covariant ``local'' scalars we obtained fluctuations and  perturbations (see section 2), which are exact, not approximated, quantities. The fluctuations and perturbations can be ``local'' when the comparison is with q--functions in a pointwise manner, or ``non--local'' if comparing local non--averaged values with the q--average assigned to a whole domain. 

The q--functions and their corresponding local perturbations have been applied successfully to examine various aspects and properties of LTB models: to construct an initial value formulation \cite{suss2002}, to examine inhomogeneous dark energy sources (quintessence and the Chaplygin gas) \cite{sussQL,suss2009}, to apply a dynamical systems approach \cite{sussDS1,sussDS2}, to examine their radial asymptotics \cite{RadAs}, the evolution of radial profiles of covariant scalars and void formation \cite{RadProfs}, to probe the application of Buchert's averaging formalism to LTB models \cite{sussBR,sussIU,suss2011} and to study the dynamics of non--spherical Szekeres models \cite{sussbol}. In the present article we extend and enhance previous work by considering also non--local perturbations and by discussing various properties of all perturbations not examined previously (their extension, amplitude, their use in Swiss cheese models and asymptotic properties). 

As a continuation of part I, we examine the q--scalars and their perturbations (local and non--local) in the framework of an initial value parametrization that is introduced in section 3, so that all relevant quantities can be scaled with respect to their values at an arbitrary  fiducial (or ``initial'') time slice. This initial value parametrization emphasizes the role of q--scalars  as LTB scalars that behave as ``effective'' FLRW scalars, as they (i) satisfy FLRW time dynamics, (ii) mimic FLRW expressions that are widely used in the literature (for example, in many of the  void models \cite{obs1,obs2,kolb,GBH,alnes,brouzakis,bismanot,swisscheese,endqvist,bisnotval,clarkson,marranot}), and (iii) identify for each domain a unique FLRW background state  through Darmois matching conditions (though an actual matching with a FLRW region is optional, not mandatory).       

Considering that (as proven in part I) local perturbations convey the deviation from FLRW geometry through the ratios of Weyl to Ricci curvature invariants and anisotropic (shear) to isotropic expansion (see equations I(42a)--I(42b) and I(43)
\footnote{We will frequently use equations derived or presented in part I. We will refer to these equations by the notation ``I(X)'', where ``X'' corresponds by the equation number in part I}
), and bearing in mind that these perturbations and their associated  q--functions completely describe all proper tensors and scalar contractions, it is natural to expect that these q--scalars and their perturbations should also yield a complete and self--consistent system of evolution equations that fully determine the dynamics of the models \cite{sussQL,suss2009,sussDS1,sussDS2,RadAs,RadProfs}. In section 4 we derive these evolution equations for local and non--local perturbations. 

The evolution equations for the q--scalars and their perturbations (local and non--local) have the structure of evolution equations for exact spherical perturbations over the FLRW background (now described in terms of the q--scalars through Darmois matching conditions).  Since q--scalars are covariant LTB objects that satisfy FLRW scaling laws and time evolution equations, while the perturbations (local and non--local) effectively convey the deviation from FLRW behavior, it is natural that a rigorous formalism of exact spherical perturbations on a FLRW background can emerge from this set of variables in which the q--scalars common to FLRW  are ``zero order'' variables and the remaining scalars (including local scalars) are ``first order'' quantities. As shown in \cite{sussQL,suss2009}, such formalism arises for the case of local perturbations. We extend in section 5 this result to non--local perturbations. 

An important distinctive characteristic of the fluctuations and perturbations (local or non--local) is their extension along the radial range: they can be either  ``confined'' ({\it i.e.} localized in a given bounded comoving domain, with or without a matching with a FLRW region as in a Swiss cheese model), or ``asymptotic'' (when the domain becomes the whole time slice). We discuss the difference between confined and asymptotic perturbations in section 6, showing (in particular) that local perturbations can always be treated asymptotically, whereas asymptotic perturbations can be non--local only for LTB models that converge to a FLRW model in the asymptotic radial range \cite{RadAs}. In this latter case, the perturbations measure the ``contrast'' of local scalars $A$ with respect to a global reference value given by the asymptotic limit of the q--average functional, which coincides with the equivalent scalar $\tilde A$ of the FLRW asymptotic state. This type of asymptotic perturbations is often used in the literature when considering perturbations in the context of LTB models (see examples and reviews of these ``contrast'' perturbations in \cite{kras2,BKHC2009}, see also the linear regime in the Appendix of \cite{zibin}).    

Since local and non--local perturbations (whether confined or asymptotic) are different objects, we illustrate this difference in section 7 by showing how they provide a different measure of the deviation from the FLRW background: local perturbations describe this deviation through the local magnitude of radial gradients of covariant scalars, whereas non--local perturbations describe it through the familiar and intuitive notion of a ``contrast'' between local  values of non--averaged scalars and reference average values assigned to a whole domain (or to a whole slice in the asymptotic limit). As a consequence, non--local perturbations are more intuitive, as the sign of their amplitudes corresponds to the familiar positive/negative sign that we associate to over/under densities. This sign is the opposite for local perturbations: over/under densities are negative/positive. 

In spite of their differences, we show in section 8 that in the linear limit both perturbations (local and non--local) yield the familiar density perturbation equation of linear theory in the comoving gauge. In section 9 we use a numerical solution of the evolution equations derived in section 4 to present an example of a cosmological void configuration that converges to an Einstein de Sitter FLRW model in the asymptotic radial range. Besides being useful to appreciate the difference between local and non--local asymptotic perturbations and their connection to the radial profiles of LTB scalars, this numerical example shows the potential utility of the q--scalars and their evolution equations for LTB model construction. We provide a summary and final discussion in section 10, while in Appendix A we prove that shell crossing singularities  necessarily occur in Swiss cheese models matching a hyperbolic LTB region with an Einstein de Sitter background.

\section{The q--average, q--scalars and their perturbations.}

LTB dust models are characterized by the metric I(1) and the field equations I(2a)--I(2b), which we repeat below for convenience:
\footnote{This section provides a quick summary of results of the preceding article (part I) that will be needed in the present article. For more detail and explanation the reader is requested to  consult part I. }
\begin{equation} \dd s^2 = -\dd t^2 + \frac{R'{}^2}{1+2E}\,\dd r^2+R^2\left(\dd\vartheta^2+\sin^2\vartheta\,\dd\varphi^2\right),\label{ltb}\end{equation}
\bse\ba  \dot R^2 = \frac{2M}{R}+2E,\label{fieldeq1}\\
 2M'= 8\pi\rho R^2R',\label{fieldeq2}\ea\ese
where $R=R(t,r)$,\, $\dot R=u^a\nabla_a R= \partial R/\partial t$,\,\, $R'=\partial R/\partial r$,\,  $E=E( r)$,\,$M=M( r)$ and $\rho=\rho(t,r)$ is the rest mass energy density (we have set $G=c=1$ and $r$ has length units). The basic covariant fluid flow scalars of the models I(7):
\bse\ba \fl \rho,\qquad \HH=\frac{\theta}{3}\,\,\hbox{(Hubble expansion)},\qquad \KK=\frac{\RR}{6}\,\,\hbox{(spatial curvature)},\label{locscals1}\\
\fl \Sigma\,\hbox{(eigenvalue of the shear tensor)},\quad  \EE\,\hbox{(eigenvalues of the electric Weyl tensor)}\label{locscals2}\ea\ese
where the rest mass density $\rho$ is given by (\ref{fieldeq2}), the expansion scalar is $\theta =\nabla_a u^a$ and $\RR$ is the Ricci scalar of the hypersurfaces $\T[t]$ orthogonal to $u^a$ (the time slices). The local scalars (\ref{locscals1})--(\ref{locscals2}) can be computed from the metric functions by means of (\ref{fieldeq2}) and I(3)--I(6), and their 1+3 evolution equations and constraints are given by the system I(8a)--I(8d) and I(9)--I(10). 

\subsection{The q--scalars.}

The LTB scalars that are common to FLRW spacetimes are (\ref{locscals1}): $\rho,\,\HH,\,\KK$, and as we showed in part I, their q--averages (defined by I(13)) in an arbitrary fixed  spherical comoving domain $\DD[r_b]$ are given by the functionals I(14)--I(16):
\begin{equation}\fl \frac{4\pi}{3}\rhoav_q[r_b]=\frac{M_b}{R_b^3},\qquad \HHav_q[r_b]=\frac{\dot R_b}{R_b},\qquad \KKav_q[r_b]=-\frac{2E_b}{R_b^2},\label{rHKave}\end{equation}
that satisfy the constraint I(17):
\begin{equation}\HHav_q^2[r_b]= \frac{8\pi}{3}\rhoav_q[r_b]-\KKav_q[r_b],\label{aveFriedman}\end{equation}
where the subindex ${}_b$ indicates evaluation at $r=r_b$. The functionals above assign the real numbers in the right hand sides of (\ref{rHKave}) for the whole domain $\DD[r_b]$. However, if we consider the q--average definition I(13) to construct real valued functions depending on a varying domain boundary, we obtain q--functions: $\rho_q,\,\HH_q,\,\KK_q$ that comply with  
\bse\ba  \frac{4\pi}{3}\rho_q=\frac{M}{R^3},\qquad \HH_q=\frac{\dot R}{R},\qquad \KK_q=-\frac{2E}{R^2},\label{rHKq1}\\
\HH_q^2= \frac{8\pi}{3}\rho_q-\KK_q,\label{rHKq2}\ea\ese 
which is formally identical to (\ref{rHKave}) and (\ref{aveFriedman}) but hold in a point--wise manner for every $r$ (see \cite{sussBR,sussIU,suss2011} and part I for more detail on the difference between the functionals $\Aav_q[r_b]$ and functions $A_q( r)$). Notice that the q--scalars (either as functions or as functionals are not coordinate `"ansatze, but fully covariant objects since $M$ and $R$ are invariant scalars in spherically symmetric spacetimes \cite{hayward} and $E$ is related to them via (\ref{fieldeq1}) (see part I for further discussion on this issue).

Since quantities that are functions of q--scalars are themselves q--scalars (see Appendix B of part I and Appendix B of \cite{sussbol}), then we can define the q--scalar I(25) $\Omega_q=U(\rho_q,\HH_q)$ or $\Omega_q=U(\rho_q,\KK_q)$ either as a functional or as a function by:
\begin{equation} \Omega_q = \frac{8\pi\rho_q}{3\HH_q^2},\qquad \Omav_q[r_b] = \frac{8\pi\rhoav_q[r_b]}{3\HHav_q^2[r_b]},\label{Omdef}\end{equation}
which is formally identical to the FLRW Omega factor.  It is straightforward to show that the q--scalars $\rho_q,\,\HH_q,\,\KK_q,\,\Omega_q$ (whether evaluated as q--functions or as functionals in fixed domains $\DD[r_b]$) satisfy the FLRW evolution laws I(27a)--I(27b). 
\footnote{For the functionals $\Aav_q[r_b]$ the derivatives involved are $\Aav\dot{}_q[r_b]$, which can be evaluated (at $r=r_b$) either directly from (\ref{rHKave}), or with the commutation rule I(22) and the forms of the local (non--averaged) scalars in (\ref{fieldeq2}), I(3), I(4) and (\ref{Omdef}). If using I(22) for computing $\HHav\dot{}_q[r_b]$ and $\langle\Omega\rangle\dot{}_q[r_b]$ we also need to use the identities I(36) and I(37) that are proved in Appendix C of part I.}, which evidently single out the q--scalars as LTB scalars that behave as FLRW scalars (in the sense that they comply with FLRW time dynamics).

%\section{Perturbations.}

\subsection{Local perturbations.}

If $A$ and $A_q=\Aav_q$ are both evaluated as real valued functions on the same arbitrary value $r$ that denotes a varying boundary of concentric domains $\DD[r]$ for $r\geq 0$, then a local perturbation follows by the pointwise evaluation comparison at each $r$ of the ratio I(29): 
\begin{equation} \Da(r) = \frac{A( r)-\Aav_q[r]}{\Aav_q[r]} =\frac{A(r)-A_q(r)}{A_q( r)},\label{Dadef}\end{equation}   
which comply (from I(23) and I(24)) with I(30) that relates the $\Da$ with  radial gradients of $A_q$ and $A$ (also valid for the $\Aav_q$): 
\begin{equation} \Da = \frac{A'_q/A_q}{3R'/R} = \frac{1}{A_q(r)\,R^3(r)}\int_0^r{A'(\bar r)\,R^3(\bar r)\,\dd\bar r},\label{Dagrad}\end{equation}
that leads, using I(23), I(24) and I(B3), to the following linear algebraic relations among the $\Da$:
\ba 
2\Dh= \Omega_q\,\Drho +\left[1-\Omega_q\right]\Dk,\label{Dhdef} \\
\DOm = \Drho-2\Dh=\left(1-\Omega_q\right)\left(\Drho-\DKK\right),\label{DOmdef}  \ea
where $\Omega_q$ is given by (\ref{Omdef}) and $\DOm$ above is consistent with $\Omega=\Omega_q(1+\DOm)$ in I(26).

\subsection{Non--local fluctuations and perturbations.}\label{contrast}

As opposed to local perturbations in which $A$ and $A_q=\Aav_q$ evaluate at the same $r$, we can define for every fixed domain $\DD[r_b]$ non--local perturbations
\begin{equation}\fl \Daav(r,r_b)=\frac{A( r)-\Aav_q[r_b]}{\Aav_q[r_b]},\qquad 0\leq r< r_b,\quad \Aav_q[r_b]\ne 0,\label{Danl}\end{equation}
that compare local values $A( r)$ inside the domain with the q--average (functional) of $A$, which is a non--local quantity assigned to the whole domain (notice that at every $\T[t]$ the value $\Aav_q[r_b]$ is effectively a constant for all $r<r_b$ and a function of $t$ for varying $\T[t]$). Evidently, the $\Daav(r,r_b)$ do not comply with (\ref{Dagrad}) and the properties that follow thereof (notice that $\partial/\partial r [\Daav(r,r_b)]=A'/\Aav_q[r_b]$). As shown in part I, the non--local fluctuations $\DENL(A)$ that give rise to non--local perturbations ($\DENL(A)=\Daav\Aav_q)$ are effectively statistical fluctuations.

\section{The q--scalars define a FLRW ``background state''.}

\subsection{The q--scalars as LTB objects that look like FLRW expressions.} 

``FLRW look alike'' expressions are often introduced in various applications of LTB models, specially in a lot of recent articles looking at LTB void models \cite{obs1,obs2,kolb,GBH,alnes,brouzakis,bismanot,swisscheese,endqvist,bisnotval,clarkson,marranot}. These expressions are introduced in these references as ``convenient'' ansatzes, without any justification other than their ``FLRW look alike'' forms, thus ignoring the fact that they can be defined rigorously as q--scalars that emerge from the weighed average I(13), and thus are fully covariant quantities related to curvature and kinematic invariants (see section 6 of part I). The  ``FLRW look alike'' expressions follow readily by  parametrizing the metric functions in (\ref{ltb}) in terms of their fiducial values at an arbitrary time slice $t=t_0$. Considering the coordinate choice $R_0 =r$
\footnote{This coordinate choice is not appropriate for LTB models whose time slices have  spherical $\mathbb{S}^3$ topology or lack symmetry centers. For such models $R'_0$ (and thus $R'$) are no longer monotonical on $r$: they must change sign at a fixed $r$ (a turning value) at every time slice. This turning value is a common zero with the gradients of all scalars \cite{RadProfs}. }
, we can transform (\ref{ltb}) into the FLRW ``look alike'' metric: 
\bse\ba \fl \dd s^2 = -\dd t^2 + a^2\,\left[\frac{\Gamma\,\dd r^2}{1-\KK_{q0}\,r^2}+r^2\left(\dd\vartheta^2+\sin^2\vartheta\dd\varphi^2\right)\right],\label{ltbbis}\\
 \fl a \equiv \frac{R}{R_0}=\frac{R}{r},\qquad \frac{\dot a}{a}=\HH_q,\qquad \Gamma \equiv \frac{R'/R}{R'_0/R_0}=\frac{rR'}{R}=1+\frac{ra'}{a},\label{aG}\ea\ese
where the relation between the scale factor $a$ and $\HH_q$ follows from (\ref{rHKq2}) (see section 7 of part I). Under the parametrization (\ref{ltbbis})--(\ref{aG}) the q--scalars $\rho_q,\,\HH_q,\,\KK_q,\,\Omega_q$ in (\ref{rHKq1})--(\ref{rHKq2}) and (\ref{Omdef}) (and their functional equivalents) take the following ``FLRW look alike'' forms that often appear in the literature:
\bse\ba \rho_q = \frac{\rho_{q0}}{a^3},\qquad \KK_q = \frac{\KK_{q0}}{a^2}\label{rk}\\
\HH_q^2 = \HH_{q0}\left[\frac{\Omega_{q0}}{a^3}+\frac{1-\Omega_{q0}}{a^2}\right],\qquad \Omega_q =\frac{\Omega_{q0}}{\Omega_{q0}+(1-\Omega_{q0})\,a},\label{HOm}\ea\ese
where the subindex ${}_0$ indicates evaluation at $t=t_0$.      

\subsection{The FLRW ``background state'' through Darmois matching conditions.} 

For every fluid flow FLRW scalar  $\tilde A=\tilde\rho,\,\tilde\HH,\,\tilde\KK,\,\tilde\Omega$ (we denote henceforth FLRW scalars by a tilde) and its ``FLRW equivalent'' LTB q--scalar $A_q=\rho_q,\,\HH_q,\,\KK_q,\,\Omega_q$, the value $A_{qb}= A_q(t,r_b)$ identifies, for each domain $\DD[r_b]$ of an LTB model, a specific FLRW dust model that could be smoothly matched (under Darmois matching conditions) at an arbitrary finite comoving radius $r=r_b$ that also marks the boundary of $\DD[r_b]$. Consider a dust FLRW universe with metric 
\begin{equation} \dd s^2 = -\dd t^2 + \tilde a^2(t)\,\left[\frac{\dd r^2}{1-k_0\,\ell_0^{-2}r^2}+r^2\left(\dd\vartheta^2+\sin^2\vartheta\,\dd\varphi^2\right)\right],\label{FLRW}\end{equation}
where $k_0=0,\pm 1$ ,\, $\ell_0$ is an arbitrary length scale and $\tilde a(t)$ is the dimensionless FLRW scale factor. Darmois conditions (necessary and sufficient) for the smoothness of the matching of (\ref{FLRW}) with an LTB model along a comoving boundary $r=r_b$ are given by \cite{ltbstuff,suss2009} 
\bse\ba \fl \rho_{qb}=\rhoav_{q}[r_b]=\tilde\rho(t)=\frac{\tilde\rho_0}{\tilde a^3},\qquad  \KK_{qb}=\KKav_{q}[r_b]=\tilde\KK(t)=\frac{\tilde\KK_0}{\tilde a^2},\label{mcon1}\\
\fl \HH_{qb}=\HHav_{q}[r_b]=\tilde\HH(t) = \frac{\dot{\tilde a}}{\tilde a},\label{mcon2}\qquad 
\Omega_{qb}=\Omav_{q}[r_b]=\tilde\Omega(t) = \frac{8\pi\tilde\rho}{3\tilde\HH^2}\\
\fl R_b = \tilde a(t)\,r\qquad \hbox{or}\quad a_b = \tilde a,\label{mcon3}\ea\ese
where the subindex ${}_0$ denotes evaluation at a fiducial hypersurface $t=t_0$ and $\tilde a(t_0)=1$ holds. We note that the continuity of the q--scalars under the matching conditions (\ref{mcon1}) and (\ref{mcon2}) is strikingly evident if we use the parametrization (\ref{ltbbis})--(\ref{aG}) and (\ref{rk})--(\ref{HOm}) with $\tilde A_0=A_{qb0}=A_q(t_0,r_b)$ holding for $A=\rho,\,\HH,\,\KK,\,\Omega$. However, from (\ref{Dagrad}) and (\ref{locscals1})--(\ref{locscals2}), it is evident that $A_b\ne A_{qb}$ and $A'_q,\,A'_q\ne 0$ hold in general, and thus the local scalars (\ref{locscals1})--(\ref{locscals2}) and the gradients $A'$ and $A'_q$ do not comply with the matching conditions (\ref{mcon1})--(\ref{mcon3}).  

It is important to remark that the continuity of the $A_q$ under Darmois matching conditions is simply a formal rigorous procedure to identify for every $\DD[r_b]$ of an LTB model a particular FLRW dust model that can be defined as a reference ``background state''.  We use the term ``state'' to emphasize that this identification does not force us to consider an actual matching with a FLRW region, which would yield a ``Swiss cheese'' configuration in which the background state becomes also an actual background spacetime. Likewise, the $A_q$ also allow us to define a ``FLRW equivalent'' region to every domain $\DD[r_b]$, as they provide through (\ref{mcon1})--(\ref{mcon3}) the values of the FLRW scalars $\tilde A$ if the whole domain was replaced by an equivalent spherical comoving section of a FLRW spacetime (without mass or volume compensation \cite{volmatch}). 

Therefore, having determined the FLRW background state, the discontinuity of the local scalars and the gradients $A'$ and $A'_q$ is not problematic if we do not wish to construct an actual Swiss cheese model through a smooth matching at the domain's boundary $r=r_b$. In this latter case we can avoid discontinuities by demanding (besides (\ref{mcon1})--(\ref{mcon3})) also the continuity of local scalars at $r=r_b$ through the following extra supplementary condition:   
\begin{equation}  A'_{qb}=0,\quad \Rightarrow \quad A_b=A_{qb} \quad\hbox{for}\quad A = \rho,\,\KK,\,\HH,\,\Omega,\label{mcon4}\end{equation}
which forces $A$ to coincide with $A_q$ at $r=r_b$, and thus explains the appearance (see panels (b) and (d) of figures 1 and 2) of ``humps'' (if $A'> 0$) or ``bags'' (if $A'< 0$) in the radial profiles of local scalars (this has been noted in the local density profiles in Swiss cheese models in LTB void models \cite{kolb,brouzakis,bismanot,swisscheese}, see reviews in chapter 5.3.5 of \cite{BKHC2009} and in \cite{marranot}). 

\section{Evolution equations.}

While the q--scalars $A_q$ or $\Aav_q[r_b]$ satisfy FLRW evolution laws, such as I(27a)--I(27b), these scalars are not fully determined by these FLRW equations because, unlike their equivalent FLRW scalars $\tilde A$, they either depend directly on $r$ or $r_b$. The missing dynamical information is provided by the evolution equations for the $\Da$ and the $\Daav$. 

\subsection{Local perturbations.}  

As shown in part I, LTB tensors and scalar invariants, as well as the local covariant scalars (\ref{locscals1})--(\ref{locscals2}) are expressible in terms of q--scalars and their local perturbations:
\bse\ba\fl  \rho = \rho_q\,\left[1+\Drho\right],\quad \HH = \HH_q\,\left[1+\Dh\right],\quad \KK=\KK_q\,\left[1+\Dk\right],\label{locscals1a}\\
\fl \Sigma = -\HH_q\,\Dh,\qquad \EE = -\frac{4\pi}{3}\rho_q\Drho,\label{locscals2a}\ea\ese
the evolution equations for the variables $\rho_q,\,\HH_q,\, \Drho,\,\Dh$ should yield a self--consistent and complete set of evolution equations for the LTB models. This system follows readily by inserting (\ref{locscals1a}) and (\ref{locscals2a}) into the ``1+3'' system I(8a)--I(8d) and its constraints I(9)--I(10). The result is the evolution equations
\bse\ba \fl\dot \rho_q &=& -3 \rho_q\HH_q,\label{EVq11}\\
\fl\dot \HH_q &=& -\HH_q^2-\frac{4\pi}{3}\rho_q, \label{EVq12}\\
\fl\dot\delta^{(\rho)} &=& -3(1+\Drho)\,\HH_q\Dh,\label{EVq13}\\
\fl\dot\delta^{(\HH)} &=& -(1+3\Dh)\,\HH_q\Dh+\frac{4\pi\,\rho_q}{3\HH_q}(\Dh-\Drho),\label{EVq14}\ea\ese
plus the algebraic constraints
\begin{equation}\fl \HH_q^2 = \frac{8\pi}{3}\rho_q - \KK_q,\quad 2\Dh= \Omega_q\,\Drho +\left(1-\Omega_q\right)\DKK,\quad \DOm = \Drho - 2\Dh,\label{qconstr}\end{equation}
that exactly coincide with the general relations (\ref{rHKq2}), (\ref{Dhdef}) and (\ref{DOmdef}), hence they hold at all $t$ ({\it i.e.} they propagate in time). The following points are worth remarking:
\begin{itemize}
\item The constraints I(9) of the 1+3 system I(8a)--I(8d) are satisfied trivially from (\ref{Dagrad}) applied to $\Drho$ and $\Dh$. 
\item The first two constraints in (\ref{qconstr}) follow from substituting (\ref{locscals2a}) into I(10) and using (\ref{fieldeq1}) and (\ref{rHKq1}), while the third one is obtained by differentiating (\ref{Omdef}) with respect to $r$ and applying (\ref{Dagrad}). Once (\ref{EVq11})--(\ref{EVq14}) is solved these constrains and (\ref{Omdef}) allow us to compute the remaining q--scalars ad their perturbations: $\Omega_q,\,\KK_q,\,\Dk,\,\DOm$. 
\item The fact that the constraints (\ref{qconstr}) of the system (\ref{EVq11})--(\ref{EVq14}) are algebraic implies a great simplification of the numeric treatment of these fluid flow evolution equations, as they can be effectively integrated as a system of autonomous ODE's in which the initial conditions are restricted by the algebraic constraints, and thus it is far easier to handle than the 1+3 system I(8a)--I(8d) in which the constraints I(9) are partial differential equations on $r$ that must be solved before the time integration of the equations. The example of the void model given in section 9 has been obtained from the numerical integration of (\ref{EVq11})--(\ref{EVq14}).
\end{itemize}
Considering the relation between the $A_q$ and $\Da$ in the constraints (\ref{qconstr}), each of the following combination of four scalars:
\ba \fl \hbox{any two of}\,\, A_q = \{\rho_q,\,\HH_q,\,\KK_q,\,\Omega_q\},\,\, 
\hbox{plus any two of}\,\, \Da = \{\Drho,\,\Dh,\,\DKK,\,\DOm\},\nonumber \ea
provides a full covariant scalar representation for the models, since the remaining pairs of $A_q$ and $\Da$ can be obtained from these algebraic constraints. Each of these representations yields a self--consistent and complete set of evolution equations that is alternative (and equivalent) to the analytic solutions of (\ref{fieldeq1}) (see Appendix A of part I) and to the numerical integration of the ``1+3'' system I(8a)--I(8d), and thus, they completely determine the dynamics of the models.

The evolution equations (\ref{EVq11})--(\ref{EVq14}) correspond to the representation $\{\rho_q,\,\HH_q,\,\Drho,\,\Dh\}$, which is useful to compare with the spherical collapse model and perturbative scenarios of structure formation that consider the density and Hubble velocity as dynamical variables. However, a more appropriate representation for cosmological applications (for example void models) is furnished by the scalars $\{\Omega_q,\,\HH_q,\,\DOm,\,\Dh\}$, leading to the following evolution equations:
\bse\ba \dot\HH_q = -(1+\frac{1}{2}\Omega_q)\,\HH_q^2,\label{Evq111}\\
\dot \Omega_q = -\Omega_q(1-\Omega_q)\,\HH_q,\label{Evq222}\\
\dot \Dh = -\left[(1+3\Dh)\Dh+\frac{1}{2}\Omega_q(\Dh+\DOm)\right]\,\HH_q,\label{Evq333}\\
\dot \DOm = -\left[(1+3\DOm)\Dh-\Omega_q(\Dh+\DOm)\right]\,\HH_q,\label{Evq444}\ea\ese 
though, as opposed to the local scalars $\rho,\,\HH,\,\KK$, the local scalar $\Omega$ defined by I(26) lacks a simple direct physical interpretation, besides being a generalization of the FLRW scalar $\tilde\Omega$.    

\subsection{Non--local perturbations.}

If we consider non--local perturbations (\ref{Danl}) in an arbitrary fixed domain $\DD[r_b]$, then only the local scalars (\ref{locscals1}) that are common to FLRW can be expressed in terms of the variables $\{\Aav_q[r_b],\,\Daav\}$ in a similar manner as in (\ref{locscals1a}) --(\ref{locscals2a}):
\begin{equation}\fl  \rho = \rhoav_q[r_b]\,\left[1+\Drhoav\right],\quad \HH = \HHav_q[r_b]\,\left[1+\Dhav\right],\quad \KK=\KKav_q[r_b]\,\left[1+\DKKav\right].\label{locscals11}\end{equation}
The remaining local scalars, $\Sigma$ and $\EE$ cannot be expressed as (\ref{locscals2}) purely in terms of non--local perturbations:
\begin{equation}\fl \Sigma( r) = \HH_q( r) -\HHav_q[r_b](1+\Dhav),\qquad \EE( r) = \frac{4\pi}{3}\left[\rho_q( r)-\rhoav_q[r_b]\Drhoav\right],\label{locscals22}\end{equation}
and as a consequence, the variables $\{\Aav_q[r_b],\,\Daav\},\,A=\rho,\,\HH,\,\KK$ do not provide a complete scalar representation of the dynamics of LTB models, which is not surprising because for any fixed domain the $\Aav_q[r_b]$ depend only on $t$ and constitute boundary conditions for the $A_q$ at $r=r_b$. From (\ref{Danl}) and (\ref{locscals11}), local and non--local perturbations are related by  
\begin{equation}1+\Daav = \frac{A_q( r)}{\Aav_q[r_b]}(1+\Da),\label{deltas}\end{equation}
which shows that they only coincide at the boundary of each $\DD[r_b]$ where $A_q(r_b)=\Aav_q[r_b]$. Since their time derivatives are evidently different, it is reasonable to expect that the evolution equations of the $\Da$ and $\Daav$ will be different. Inserting (\ref{deltas}) for $A=\rho,\,\HH$ into (\ref{EVq13})--(\ref{EVq14}) yields these equations:
\bse\ba 
\fl\dot\Drhoav &=& -3(1+\Drhoav)\,\HHav_q[r_b]\,\Dhav,\label{EVq31}\\
\fl\dot\Dhav &=& -(1+3\Dhav)\,\HHav_q[r_b]\,\Dhav+\frac{4\pi\,\rhoav_q[r_b]}{3\HHav_q^2[r_b]}(\Dhav-\Drhoav)\nonumber\\
\fl &{}&-2\HHav_q[r_b]\left(1-\frac{\HH_q( r)}{\HHav_q[r_b]}\right)^2 +4(\HH_q( r)-\HHav_q[r_b])\Dhav,\label{EVq32}\ea\ese
which, in order to render a fully complete system to describe the dynamics of the models in an arbitrary fixed domain $\DD[r_b]$, needs to be supplemented by the evolution equations for $\rhoav_q[r_b],\,\HHav_q[r_b]$ and $\rho_q,\,\HH_q$:
\bse\ba
\fl \HHav\dot{}_q[r_b] = -\HHav_q^2[r_b]-\frac{4\pi}{3}\rhoav_q[r_b],\qquad \dot\HH_q( r) = -\HH_q^2( r)-\frac{4\pi}{3}\rho_q( r),\label{EVq33}\\
\fl \rhoav\dot{}_q[r_b] = -3\HHav_q[r_b]\rhoav_q[r_0],\qquad  \dot\rho_q( r) = -3\HH_q( r)\rho_q( r),\label{EVq34} 
\ea\ese
while the constraints take the form (\ref{qconstr}) and with the $\Da$ expressed in terms of the $\Daav$ by (\ref{deltas}) (there is an extra constraint given by (\ref{aveFriedman})). It is interesting to remark that (\ref{EVq31}) is identical to (\ref{EVq13}), but (\ref{EVq32}) differs from (\ref{EVq13}) by the terms with $\HH_q$, which explains the need to add the evolution equations for $\rho_q$ and $\HH_q$. Notice that the non--local evolution equations become identical to the local ones (\ref{EVq11})--(\ref{EVq14}) at the domain boundary in the limit $r\to r_b$ where $\HH_q(r_b)=\HHav_q[r_b]$. Also the non--local evolution equations depend on averages $\rho_q( r)=\rhoav_q[r],\,\HH_q( r)=\HHav_q[r]$ for inner points of $\DD[r_b]$ ($r<r_b$) and are considerably more complicated than (\ref{EVq11})--(\ref{EVq14}). 

The self--consistency of the system (\ref{EVq31})--(\ref{EVq34}) can be proved easily by comparing the mixed derivatives $[\dot\Daav]'$ obtained from $[\Daav]'=A'_q( r)/\Aav_q[r_b]$ with those that follow from the radial derivative of the right hand sides of (\ref{EVq31})--(\ref{EVq32}). Evolution equations for non--local perturbations on the representation $\{\Omega_q,\,\HH_q,\,\DOm,\,\Dh\}$ can also be constructed, but the resulting equations are more cumbersome than (\ref{EVq32}) and (\ref{EVq34}). However, it always possible (and easier) to solve the evolution equations for local perturbations (either (\ref{EVq11})--((\ref{EVq14}) or (\ref{Evq111})--((\ref{Evq444})) and then compute the non--local perturbations through the relation (\ref{deltas}) (this is what was done in section 9). 

\subsection{Evolution equations without back--reaction.}

The evolution equations constructed with q--scalars and their perturbations (local and non--local) lack the back--reaction correlation terms that appear in the evolution equations that follow from Buchert's formalism (notice that (\ref{EVq12}) is identical to I(52) with $\QQ_q=0$) in I(53)). Also, these equations (in any q--scalar representation) form complete and self--consistent systems that can be integrated without further assumptions for any given set of consistent initial conditions. On the other hand, in order to close and integrate Buchert's evolution equations it is necessary to make specific assumptions linking the back--reaction terms with the averaged scalars \cite{buchert,buchCQG,buchstruct}.   

\section{A formalism of exact perturbations on a FLRW background.}\label{locpert}

It is evident that (as pointed out in previous work \cite{sussQL,suss2009}) the system (\ref{EVq11})--((\ref{EVq14}) has the structure of evolution equations of spherical dust perturbations (the $\Da$) on a FLRW background (defined by the $A_q$). While the non--local perturbations were not considered in these references, the same resemblance to dust perturbation holds for systems like (\ref{EVq31})--(\ref{EVq32}) and (\ref{EVq33})--(\ref{EVq34}) involving $\Aav_q[r_b],\,A_q( r)$ and $\Daav(r,r_b)$. 

Following the standard methodology \cite{zibin,ellisbruni89,BDE,1plus3,dunsbyetal,LRS,pertmap1,pertmap2}, a perturbation formalism linking a ``lumpy'' spacetime (LTB model) and a homogeneous ``background spacetime'' (a FLRW dust model) in a domain $\DD[r]$ can be defined by comparing local LTB variables with LTB objects that can define an FLRW ``background state'' of ``zero order'' variables by means of suitable maps (evidently, these objects are the q--scalars). However, such maps must also deal with gauge issues involving a specific time slicing and coordinates. Since the boundary of every spherical comoving FLRW region can be mapped into the boundary of a domain $\DD[r]$ of an LTB model by the  Darmois matching conditions (\ref{mcon1})--(\ref{mcon2}), and bearing in mind that LTB models and dust FLRW universes are both (i) spherically symmetric, (ii) have a geodesic 4--velocity and (iii) their full dynamics reduces to scalar modes~\cite{zibin,LRS}, a perturbation formalism associated with the q--scalars in domains $\DD[r]$ can be defined rigorously by means of maps between FLRW covariant scalars and the q--scalars with all gauge issues resolved.

\subsection{The perturbation maps.} 

Let $\MM$ be an LTB model and $X(\DD[r])$ the set of all covariant scalars in an arbitrary domain $\DD[r]$ of $\MM$. Let $\tilde \MM$ be a dust FLRW model and  $\tilde X(\tilde\MM)$ the set of covariant scalars of $\tilde \MM$:  
\begin{quote}
{\bf The local perturbation map.}  For every $\DD[r]$ in $\MM$ there exists a model $\tilde \MM$ such that $A_q(t,r)=\tilde A(t)$ holds for every $\tilde A\in \tilde X(\tilde \MM)$ and $A\in  X(\MM)$ and for all $t$. The following maps
\begin{equation} \fl \Phi:  \tilde X(\tilde\MM) \to X(\DD[r]),\qquad \tilde A  \mapsto \Phi(\tilde A)=A_q(r) \in X(\DD[r]),\label{map1}\end{equation}
\ba \fl \delta^{(\hskip 0.1cm)}: X(\DD[r])\to X(\DD[r]),\quad  A\mapsto \Da = \frac{A-\Phi(\tilde A)}{\Phi(\tilde A)}=\frac{A-A_q}{A_q}=\frac{A-\Aav_q}{\Aav_q},\nonumber\\ \label{map2}\ea
define for every $\DD[r]$ a ``background state'' associated with an FLRW cosmology $\tilde\MM$ and  local exact perturbations of scalars $A$ obtained by comparing them with the LTB scalars produced by the map $\Phi$.
\end{quote}
An analogous scalar perturbation formalism for the non--local fluctuations can be defined along the lines of (\ref{map1}) and (\ref{map2}):
\begin{quote}
{\bf The non--local perturbation map.} Let $Y(\DD[r_b])$ be the set of all linear functionals in an arbitrary fixed comoving domain $\DD[r_b]\subset \T[t]$.  For every FLRW covariant scalar $\tilde A\in \tilde X(\tilde \MM)$ and every $\DD[r_b]$ the following map 
\begin{equation} \PhiNL:\tilde X(\tilde \MM) \to Y(\DD[r_b]),\qquad \tilde A\mapsto \Aav_q[r_b],\label{map1a}\end{equation}
defines a ``background state'' associated with a FLRW cosmology $\tilde\MM$ but consisting of the functionals I(13). The definition of the non--local perturbation of the scalar $A$ is then the map $\delta_{\textrm{\tiny{NL}}}^{(\hskip 0.1cm)}: \tilde X(\tilde\MM) \times X(\DD[r_b])\to X(\DD[r_0]) \times Y(\DD[r_0])$ such that
\begin{equation} (\tilde A,A) \mapsto \Daav = \frac{A-\PhiNL(\tilde A)}{\PhiNL(\tilde A)}=\frac{A( r)-\Aav_q[r_b]}{\Aav_q[r_b]}, \label{map2a}\end{equation}
which provides a pointwise comparison between the real valued functions $A$ and their associated functionals $\Aav_q[r_b]$ assigned to the whole domain $\DD[r_b]$.
\end{quote}
It is important to emphasize that in the definitions above we distinguish between the FLRW ``background state'' and the FLRW ``background spacetime'' $\tilde \MM$. As commented in section 3, the former is defined by the q--scalars:\, $\{\rho_q,\,\HH_q\,\KK_q,\,\Omega_q\}$ or $\{\rhoav_q[r_b],\,\HHav_q[r_b],\,\KKav_q[r_b],\,\Omav_q[r_b]\}$, which satisfy FLRW dynamics and relate to the FLRW scalars $\{\tilde\rho,\,\tilde\HH,\,\tilde\KK,\,\tilde\Omega\}$ of the background spacetime $\tilde\MM$ by continuity under the Darmois matching conditions at an arbitrary constant $r$. These q--scalars are (in the perturbation maps) the ``zero order'' variables, with the  ``first order variables'' being the local fluid flow scalars $A$ related to the former by the perturbations $\Da$ or $\Daav$. 

Notice that the perturbation formalism defined by (\ref{map1}) and (\ref{map2}) is covariant because the background variables and the perturbations (zero and first order variables) are coordinate independent LTB objects (see section 6 of part I). In fact, following the Stewart--Walker lemma \cite{ellisbruni89,BDE,1plus3,SWlemma}, the perturbations $\Da$ and $\Daav$ are also gauge invariant (even in the usual sense of \cite{bardeen}) as they vanish in the FLRW spacetime $\tilde\MM$ associated with the background state through (\ref{map1}) and (\ref{map2}).
\begin{figure}
\begin{center}
\includegraphics[scale=0.3]{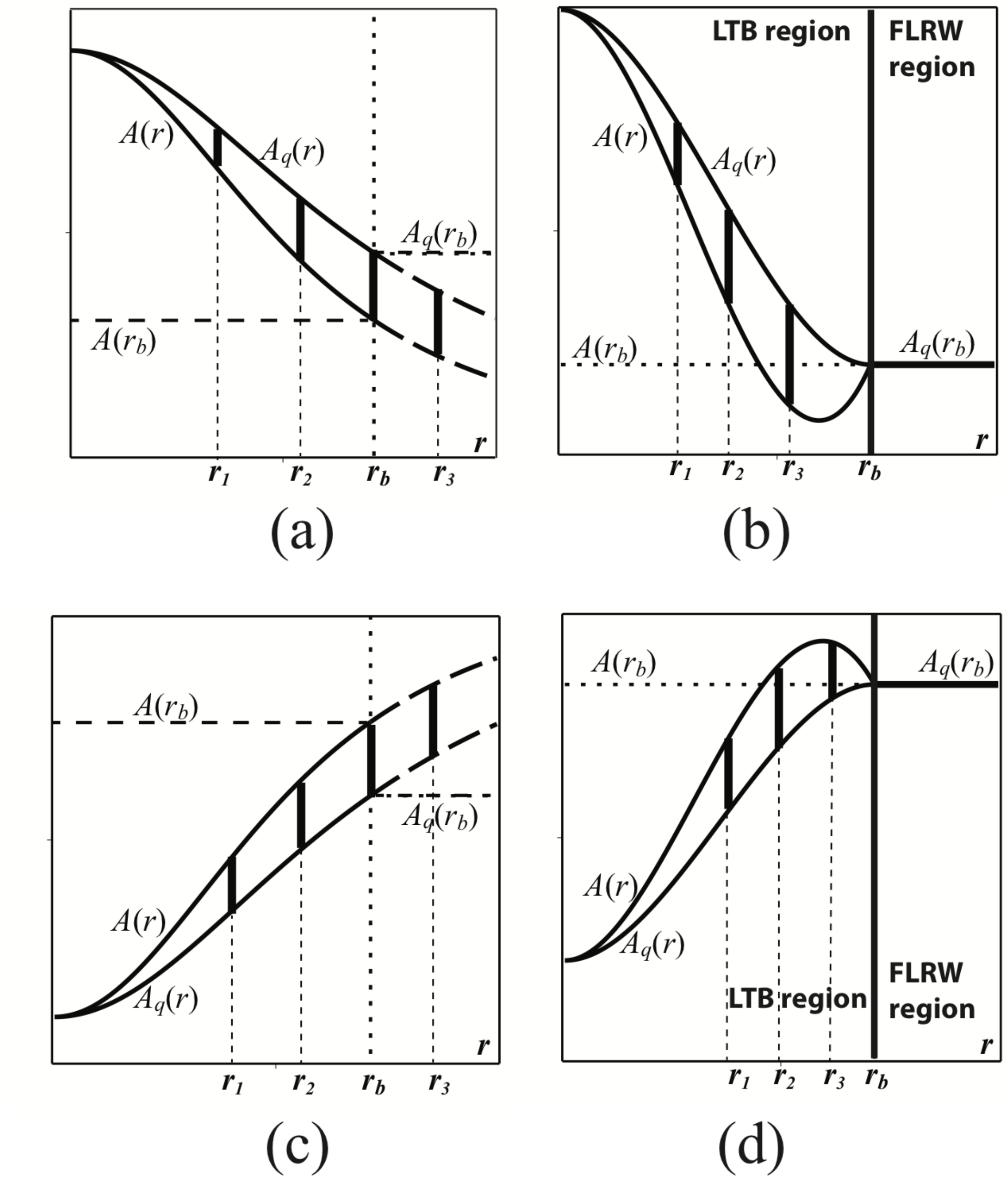}
\caption{{\bf Local confined fluctuations.} The fluctuations $\Del(A)=A-A_q=A_q\Da$ (thick vertical lines) follow by pointwise comparison of the scalars $A$ and $A_q$ along $r$ in an arbitrary time slice $\T[t]$. Clump and void profiles are displayed in panels (a)--(b) and ( c)--(d), the case of a Swiss cheese model matched to a FLRW region at $r=r_b$ is displayed in panels (b) and (d). Notice how in these latter cases $\Del(A)=0$ at $r=r_b$, thus forcing $A_q(r_b)=A(r_b)$, which explains the ``humps'' and ``bags'' in the radial profile of $A$. Shell crossing singularities necessarily emerge in Swiss cheese models in which the LTB region is hyperbolic and the FLRW region is Einstein de Sitter (see Appendix A).}
\label{fig1}
\end{center}
\end{figure}
\begin{figure}
\begin{center}
\includegraphics[scale=0.3]{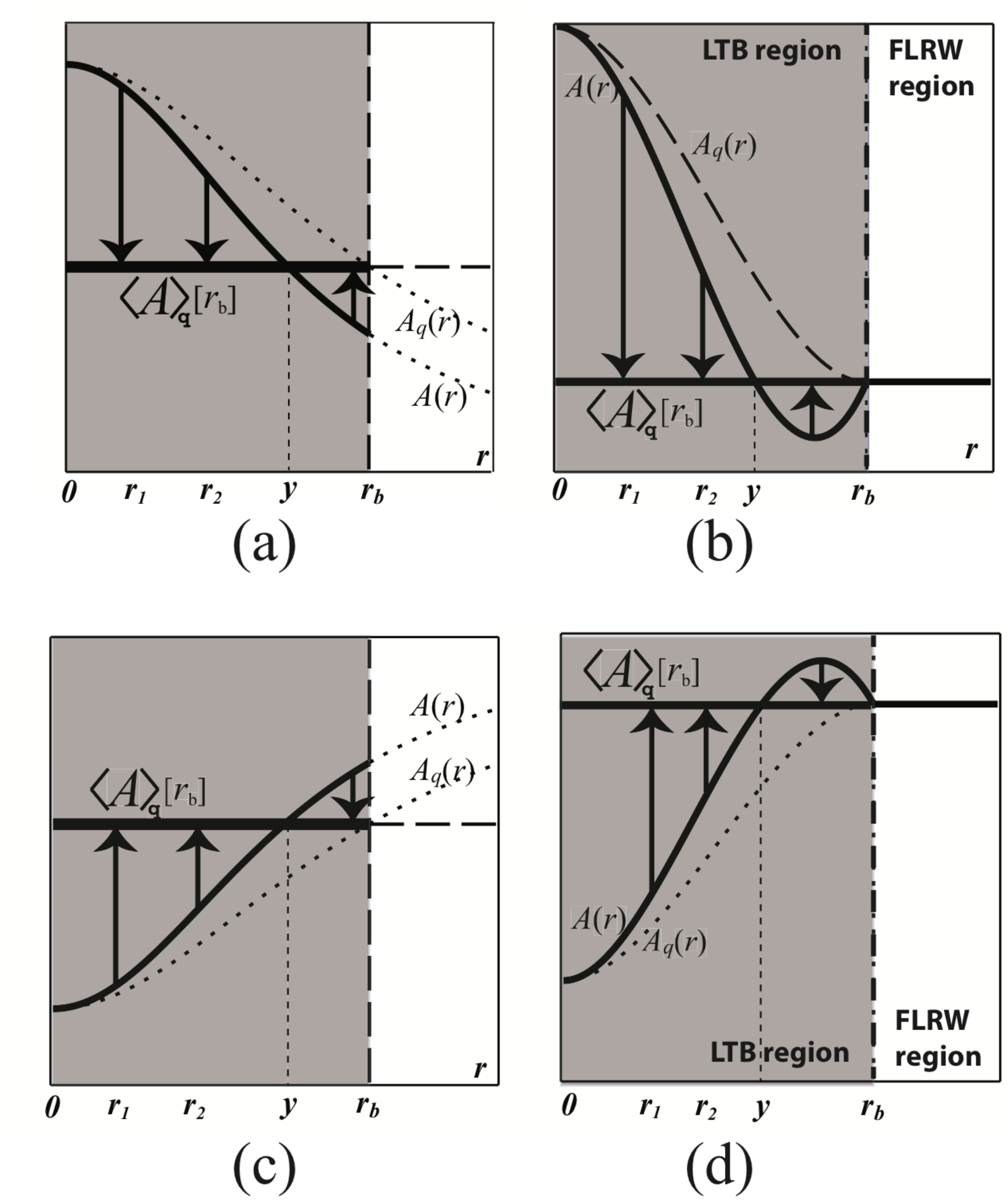}
\caption{{\bf Non--local confined fluctuations.} The panels display (as in figure 1) clump and void profiles with and without  a Swiss cheese matching to a FLRW region at $r=r_b$ in an arbitrary time slice $\T[t]$. The non--local fluctuations $\DENL(A)=A-\Aav_q=\Aav_q\Daav$ (the arrows) compare local values of $A( r)$ with the average $\Aav_q[r_b]$ (thick line) assigned to the whole domain $\DD[r_b]$ (shaded region). Notice (as in figure 1) that $\DENL(A)$ only vanishes at $r=r_b$ if there is a Swiss cheese matching (panels (b) and (d)).}
\label{fig2}
\end{center}
\end{figure}

\subsection{LTB models as exact perturbations.}

As opposed to the conventional approach to perturbations, either the traditional one with gauge invariant variables  \cite{pertmap1,pertmap2,bardeen} or the covariant formalism of Ellis {\it et al} \cite{ellisbruni89,BDE,1plus3}), the perturbations that emerge from (\ref{map1})--(\ref{map2}) and (\ref{map1a})--(\ref{map2a}) are exact (not approximate) quantities, and thus do not lead to some unknown ``near FLRW'' space-time on the basis of a linearization process (though a rigorous linear limit can be defined, see section 8). Instead, the $\{A_q,\,\Da\}$ or the $\{\Aav_q[r_0],\,\Daav\}$ with $A=\rho,\,\HH,\,\Omega,\,\KK$ express a known class of spacetimes (generic LTB models) as exact spherical perturbations on an abstract FLRW background state defined by LTB objects: the $A_q$ along all domains $\DD[r]$ with varying boundary or the $\Aav_q[r_b]$ at a fixed domain. Evidently, these perturbation formalisms are special in the sense that they preserve the spherical symmetry and the dust source of the FLRW background. However, these formalisms are applicable to any space-time compatible with an LTB metric in the comoving frame and having an anisotropic fluid source \cite{suss2009}, and can be readily generalized for the non--spherical Szekeres dust models \cite{sussbol}.

\section{Confined perturbations, Swiss cheese models and asymptotic perturbations.}

In the definitions (\ref{map1})--(\ref{map2}) and (\ref{map1a})--(\ref{map2a}) we assumed domains $\DD[r]$ or $\DD[r_b]$ with $r$ and $r_b$ finite. As we show below, the resulting perturbations can also be considered in the asymptotic case when $r_b\to \infty$.

\subsection{Perturbations confined in comoving domains.}

As long as $r$ and $r_b$ are finite, we are considering perturbations that are confined in arbitrary bounded comoving domains of an LTB model. Depending on whether we perform a smooth match with a FLRW spacetime or not we have the following possibilities:
\begin{description}
\item[There is no matching with FLRW] (see panels (a) and ( c) of figures 1, 2 and 3). The perturbation describes the dynamics of the domain $\DD[r]$ or $\DD[r_b]$ with respect to a FLRW background state defined by (\ref{map1}) or (\ref{map1a}), which is a fictitious reference FLRW dust model $\tilde \MM$ (the background spacetime) whose scalars $\tilde A$ match (via Darmois matching conditions) the zero order variables $A_q( r)$ or $\Aav_q[r_b]$. Notice that:
\begin{itemize}
\item The fictitious reference FLRW dust model $\tilde \MM$ is necessarily different for different domains.
\item First order quantities like $A$ and the gradients $A'$ and $A'_q$ ($A=\rho,\,\HH,\,\KK,\,\Omega$) are not continuous at the domain boundaries $r$ or $r_b$. See section 3 and panels (a) and ( c) of figures 1, 2 and 3.   
\end{itemize} 
\item {\bf Swiss cheese holes smoothly matched to a FLRW region} (see panels (b) and (d) of figures 1, 2 and 3). The perturbation describes the dynamics of a comoving domain $\DD[r_b]$ with respect to a background state that in this case corresponds to the actual (non--fictitious) FLRW background spacetime matched at $r=r_b$ (under conditions (\ref{mcon1})--(\ref{mcon3})) and extending for $r>r_b$. Notice that:
\begin{itemize}
\item The resulting spacetime is a compound Swiss cheese configuration consisting of an LTB section (confined in $\DD[r_b]$) described by zero order variables $A_q$ and their perturbations $\Da$ or $\Daav$, and the FLRW background spacetime that extends for $r>r_b$. The FLRW dust model $\tilde \MM$ is necessarily different for different domains $\DD[r_b]$.
\item Darmois matching conditions only require the zero order quantities $A_q$ and $\Aav_q[r_b]$ to be continuous at $r_b$, though it is always possible to demand (\ref{mcon4}) (as extra conditions) so that local scalars $A$ and the gradients $A'$ and $A'_q$ (which are first order quantities related to $\Da$ via (\ref{Dagrad})) are also continuous at $r_b$. These extra conditions imply that the fluctuations and perturbations themselves vanish at $r_b$ (see section 3 and panels b and d of figures 1, 2 and 3).
\item Swiss cheese models have been considered in the literature in the context of fitting observations without resorting to dark energy \cite{kolb,brouzakis,bismanot,swisscheese} (see review in chapter 5.3.5 of \cite{BKHC2009} and in \cite{marranot}). However, a shell crossing singularity necessarily emerge in Swiss cheese models in which a hyperbolic LTB region is matched to an Einstein de Sitter background at finite $r_b$ (see proof in Appendix A).  
\end{itemize} 
\end{description}

\subsection{Asymptotic perturbations.}

The application of the perturbation maps (\ref{map1})--(\ref{map2}) and (\ref{map1a})--(\ref{map2a}) to the asymptotic limit $r\to\infty$ or $r_b\to\infty$ depends on the convergence of LTB scalars and their associated q--scalars along radial rays ({\it i.e.} curves $x^a = \ell\delta^a_r$, where $\ell=\int{g_{rr}\dd r}$ is the proper radial length), which are spacelike geodesics of the metric (\ref{ltb}) and of the slices $\T[t]$ with metric $h_{ab}$. As shown in \cite{RadAs}, $\ell(r)$ is a positive monotonic function in regular LTB models, and thus the asymptotic limit $r\to\infty$ corresponds to $\ell\to\infty$. 

The radial asymptotic behavior of LTB models is determined by the following asymptotic limits for the scalars $A=\rho,\,\HH,\,\KK,\,\Omega$: 
\footnote{LTB models can only converge in the asymptotic radial regime to FLRW dust models with zero or negative spatial curvature, not positive. Models convergent to Minkowski can converge to a section of Minkowski parametrized by Milne coordinates or by non--standard curvilinear coordinates.  In the Milne case $\rho_q,\,\Omega_q\to 0$ but $\HH_q,\,\KK_q$ tend to nonzero values (see \cite{RadAs}).}
\begin{equation}\fl \mathop {\lim }\limits_{r \to \infty } A(r) = \mathop {\lim }\limits_{r \to \infty } A_q(r) =\mathop {\lim }\limits_{r_b \to \infty } \Aav_q[r_b] =\left\{ \begin{array}{l}
 A_{_\infty}(t)\ne 0,\quad \hbox{Asymptotically FLRW}\\ 
  \\ 
0,\qquad \hbox{Asymptotically Minkowski}.\\ 
 \end{array} \right.\label{asympt}\end{equation}
These limits correspond to q--averages for increasingly large domains up to the situation in which $\DD[r]$ or $\DD[r_b]$ become the whole time slice $\T[t]$, and thus, we can speak of the q--average of a whole LTB model (instead of the q--average of confined domains of an LTB model). We look at the perturbations for asymptotically FLRW and Minkowski models separately below. 
\begin{itemize}
\item {\bf Asymptotically FLRW models} (see figure 4). As a consequence of (\ref{map2}) and (\ref{asympt}) local perturbations vanish asymptotically in these models:
\begin{equation} \mathop {\lim }\limits_{r \to \infty } \Da( r) =0.\label{asdeltaloc}\end{equation}
However, (\ref{map2a}) and (\ref{asympt}) lead for non--local perturbations to the following non--trivial nonzero asymptotic limit:
\begin{equation} \fl \mathop {\lim }\limits_{r_b \to \infty } \Daav(r,r_b) = \frac{A( r)-A_{_\infty}(t)}{A_{_\infty}(t)} =\frac{A( r)-\tilde A(t)}{\tilde A(t)},\qquad r\;\;\hbox{finite},\label{asdeltaNL}\end{equation}
which keeps their interpretation of a ``contrast'' between local values $A( r)$ and the domain average, with the difference that the average now corresponds to a domain that encompasses the whole slice $\T[t]$ and coincides with the corresponding scalars $A_{_\infty}(t)=\tilde A(t)$ of an asymptotic FLRW background state  defined by (\ref{map1a}). As a consequence, we can rigorously state that the q--average of every LTB model of this class is exactly the FLRW background spacetime $\tilde\MM$. Asymptotic fluctuations and perturbations for a model converging to FLRW are illustrated schematically by figure 4 and for the numerical example of section 9 by figures 6, 8, 9, 11 and 12b.

\item {\bf Asymptotically Minkowski models}. By looking at (\ref{map2a}) it is evident that non--local perturbations $\Daav$ cannot be defined in the asymptotic limit for these LTB models, since in general $A( r)\ne 0$ for finite $r$ and thus the ratio $A/\Aav_q[r_b]$ diverges as $\Aav_q[r_b]\to 0$ in the limit $r_b\to\infty$. However, local perturbations $\Da$ defined by (\ref{map2}) are not affected by the radial convergence to a Minkowski vacuum because, as shown in \cite{RadAs}, both $A\to 0$ and $A_q\to 0$ hold as $r\to\infty$ but their ratio $A/A_q$ is finite in this limit, and thus the $\Da$ in (\ref{map2}) are well defined. The only difference with asymptotically FLRW models is that the $\Da$ tend (in general) to nonzero constant values as $r\to\infty$ \cite{RadAs}. This applies also to the perturbations of the spatial curvature $\DKK$ in elliptic and hyperbolic models radially converging asymptotically to a spatially flat  Einstein de Sitter model (see the profile of the perturbation $\DKK$ in figure 12b). 
\end{itemize}

\section{Comparison between local and non--local perturbations.} 

Since the perturbations $\Da$ and the $\Daav$ are different objects, they  provide a different measure of the inhomogeneity or ``deviation'' of an LTB model from a FLRW Universe:

\begin{itemize}  
\item The $\Da$ measure this deviation through local comparison of $A( r)$ with the characteristic FLRW scalar $A_q( r)=\Aav_q[r]$ at every $r$, and is related (via (\ref{Dagrad})) to radial gradients of $A_q$ and $A$ and (via I(42a)--I(42b) and I(43)) to the ratio of Weyl to Ricci curvature and anisotropic to isotropic expansion. See figures 1 and 3. 
 
\item The $\Daav$ measure the deviation from FLRW though the notion of the ``contrast'' between local values of $A( r)$ inside an arbitrary but  fixed comoving domain ($0\leq r<r_b$) and a value $\Aav_q[r_b]$ that can be associated to the equivalent FLRW scalar $\tilde A(t)=\Aav_q[r_b]$ that characterizes the whole domain. In the asymptotic limit $r_b\to\infty$ so that $\tilde A(t)=A_{_\infty}(t)$, these perturbations provide the contrast with respect to the FLRW spacetime to which the LTB model converges asymptotically. See figures 4, 6b, 8b, 10b and 11b. The exceptional case is with spatial curvature perturbations when this FLRW spacetime is Einstein de Sitter, in which case the non--local perturbation $\DKKav$ cannot be defined, see section 9). 
\end{itemize} 
We examine below how these differences relate to the shape of the radial profile of a scalar at an arbitrary $\T[t]$ and the signs and magnitudes (``amplitude'') of the perturbations. 

\subsection{Signs of the perturbations vs type of radial profiles.}

Consider a scalar $A$ whose radial profile is monotonic throughout a given domain 
\footnote{Such domains always exist for $r_b$ finite, even if $A$ is not monotonic for values $r>r_b$. The nature and evolution of radial profiles of LTB scalars were examined extensively in reference \cite{RadProfs}. In general, density void profiles are fully compatible with regularity conditions (absence of shell crossings) for hyperbolic models, but not for elliptic or parabolic ones. Also, for models with a non--simultaneous big bang ($\tbb'\ne 0$, nonzero decaying modes) density void profiles always emerge from initial clump profiles after a transition (``profile inversion'', see figures 6a, 6b and 7a), whereas a density void profile exists for all the time evolution only in models with a simultaneous bang (zero decaying modes). A profile inversion of the expansion scalar $\HH$ necessarily occurs in elliptic models but not in hyperbolic ones (see \cite{RadProfs} for further detail). }.
Then, following I(23), the clump/void profiles are defined by
\bse\ba \fl \hbox{Clump profile}\qquad A'\leq 0\quad \Rightarrow \quad A'_q\leq 0 \quad \Rightarrow \quad A\leq A_q,\label{clump}\\
\fl \hbox{Void profile}\qquad\quad A'\geq 0\quad \Rightarrow\quad A'_q\geq 0 \quad \Rightarrow \quad A\geq A_q,\label{void}
\ea\ese
where $A(0)=A_q(0)=A_c$ and $A'(0)=A'_q(0)=0$ and we have assumed absence of shell crossings and singular layers (hence $R'>0$ holds everywhere or it has a common same order zero with $A'$ and $A'_q$ \cite{RadAs,RadProfs}). The relation between the signs of the perturbations and the profile type follows from (\ref{Dagrad}) (it can also be seen schematically in figures 1--4 and in the graphs of the profiles and perturbations of figures 6--11 of the numerical example of section 6):
\begin{itemize}
\item {\bf Local perturbations}:
\ba \fl \hbox{sign}(\Da) = \hbox{sign}(A'_q) \quad\Rightarrow\quad 
\left\{ \begin{array}{l}
 \hbox{clump profile}:\,\Da\leq 0,\\ 
  \\ 
\hbox{void profile}:\quad\Da\geq 0.\\ 
 \end{array} \right.\label{signDa}\ea   
Figure 4 depicts this sign relation. It can also be appreciated by comparing the profiles of $\rho$ and $\HH$ in figures 7 and 10 with the corresponding local perturbations $\Drho$ and $\Dh$ in figures 6a, 8a, 9a and 11a.  
\item {\bf Non--local perturbations}:

{\underline{Finite domains $\DD[r_b]$ and Swiss cheese holes}}. The sign relations in (\ref{clump}) and (\ref{void}) imply that there always exists a value $\bar r=y<r_b$ such that $A(y)=A_q(r_b)=\Aav_q[r_b]$, and thus we have from (\ref{Danl}): $\Daav(y,r_b)=0$. Since $A_c>\Aav_q[r_b]$ holds in clump profiles and $A_c<\Aav_q[r_b]$ in void profiles, then $\Daav$ changes sign as follows (see panels (b) and (d) of figures 2 and 3):
\bse\ba \fl \hbox{Clump:} \qquad \Daav > 0\;\;\hbox{for}\;\; 0\leq \bar r < y,\qquad \Daav < 0\;\;\hbox{for}\;\; y<\bar r <r_b,\label{signDanl11}\\
\fl \hbox{Void:} \qquad\quad \Daav < 0\;\;\hbox{for}\;\; 0\leq \bar r <y,\qquad \Daav > 0\;\;\hbox{for}\;\; y<\bar r <r_b,\label{signDanl12}
\ea\ese
{\underline{Asymptotic background}}  In the limit $r_b\to\infty$ we have $A\to A_{_\infty},\,\Aav_q[r_b]\to A_{_\infty}$ and $\DD[r_b]\to \T[t]$, hence for all $r$:
\bse\ba \fl \hbox{Clump:}\qquad  A> A_{_\infty}\qquad \Daav > 0, \label{signDanl21}\\
\fl \hbox{Void:} \qquad\quad A< A_{_\infty} \qquad \Daav < 0,\label{signDanl22}
\ea\ese
The relation between the type of profile and the sign of the perturbations is illustrated schematically by figure 4. It also emerges by comparing the profiles of $\rho$ and $\HH$ in figures 4, 7 and 10 with the signs of the local and non--local perturbations $\Drhoav$ and $\Dhav$ in figures 6b, 8b, 9b and 11b.
\end{itemize}  
Comparing the sign relations in (\ref{signDa}) with those in (\ref{signDanl11})--(\ref{signDanl12}) and (\ref{signDanl21})--(\ref{signDanl22}) (and looking at figures 6b, 7a and 10b) clearly indicates that non--local density perturbations $\Drhoav$ have the expected intuitive signs that we tend to infer for perturbations through the familiar notion of a ``density contrast'' with respect to a fiducial (or background) FLRW density value $\tilde\rho$: an over--density (clump) is a positive perturbation because local density is larger than $\tilde\rho$, while an under--density (void) is a negative perturbation because $\tilde\rho$ is smaller than local density, which we can now identify with $\rhoav_q[r_b]$ or its asymptotic limit $\rho_{_\infty}$ (see figure 6b). In fact, this intuitive notion of a ``contrast'' can be applied to all covariant scalars, not just the density (it is applied to the Hubble scalar $\HH$ in the numerical example in figures 8b and 11b).  On the other hand, the ``gradient'' based local perturbations $\Da$ have the opposite sign to the intuitively expected one and thus are more abstract and counter--intuitive (compare the profiles of $\rho,\,\HH$ in figures 7 and 10 with the signs of $\Drho,\,\Dh$ in figures  6a, 8a, 9a and 11a).  
\begin{figure}
\begin{center}
\includegraphics[scale=0.3]{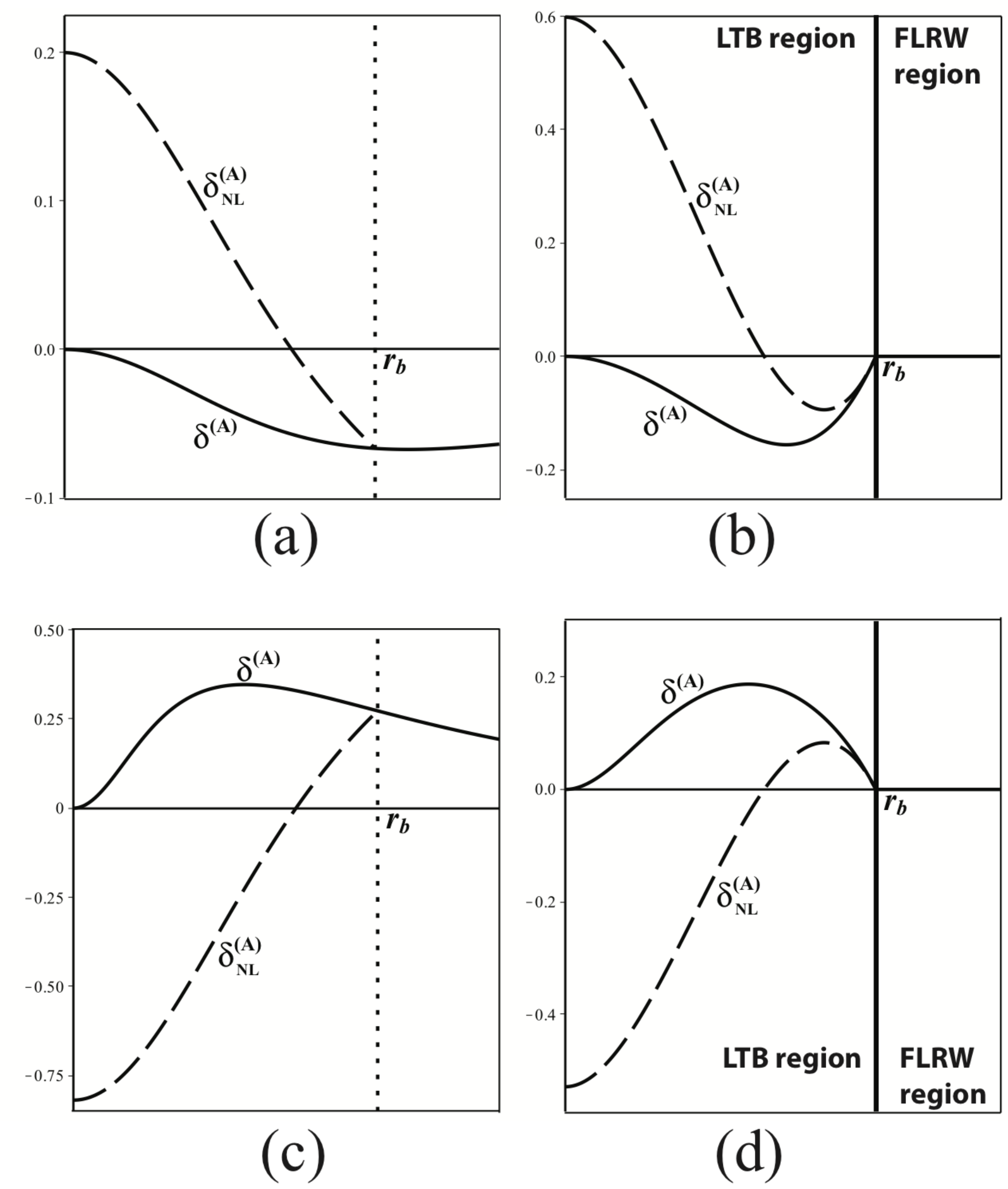}
\caption{{\bf Confined perturbations.} The panels display the local and non--local perturbations ($\Da$ and $\Daav$) for the radial profiles displayed in figures 1 and 2. Notice that both types of perturbations vanish at $r=r_b$ when there is a Swiss cheese matching (panels (b) and (d)). Notice also that $\Da$ and $\Daav$ have opposite signs around the center.}
\label{fig3}
\end{center}
\end{figure}
\begin{figure}
\begin{center}
\includegraphics[scale=0.3]{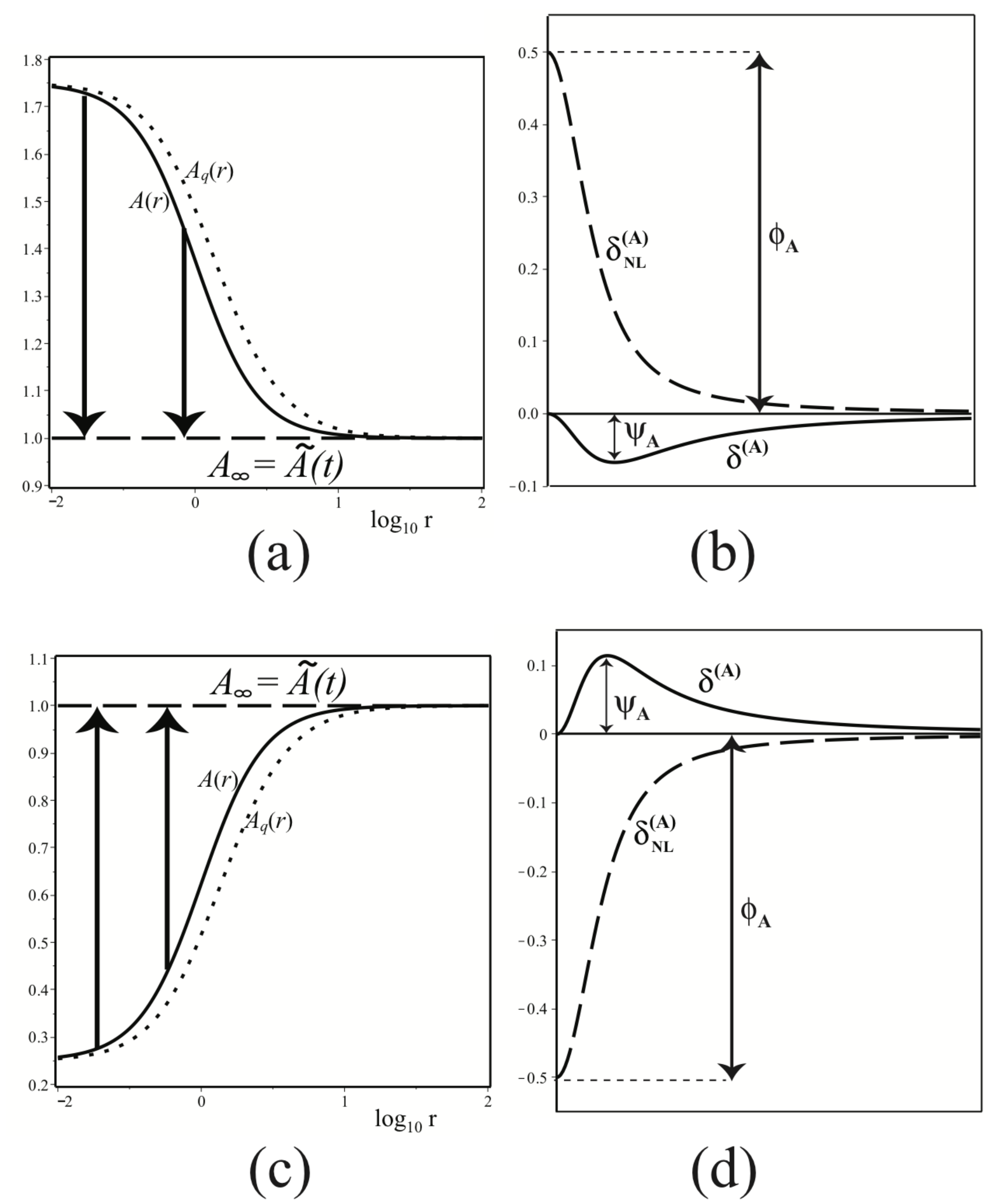}
\caption{{\bf Asymptotic fluctuations and perturbations.} Panels (a) and (b) correspond to a clump profile, while ( c) and (d) depict a void profile. Since the domain encompasses a whole time slice $\T[t]$, non--local fluctuations compare local values of $A( r)$ with the global average $\Aav_q[r_b]\to A_{_\infty}$ in the limit $r_b\to\infty$, which is equal to the corresponding scalar $\tilde A(t)$ of the FLRW asymptotic state. Non--local perturbations $\Daav$ yield the ``contrast'' between local values of $A$ and this average. The amplitude of the perturbation is $\phi_A$, while $\psi_A$ depicts the maximal value of the local perturbation $\Da$. Notice that $\Daav$ is positive/negative for the clump/void, while $\Da$ has opposite signs. Also, both perturbations vanish as $r\to\infty$. The graphs of asymptotic perturbations displayed in figures 6, 8, 9 and 11 corroborate the qualitative forms shown in panels (b) and (d).}
\label{fig4}
\end{center}
\end{figure}
\begin{figure}
\begin{center}
\includegraphics[scale=0.35]{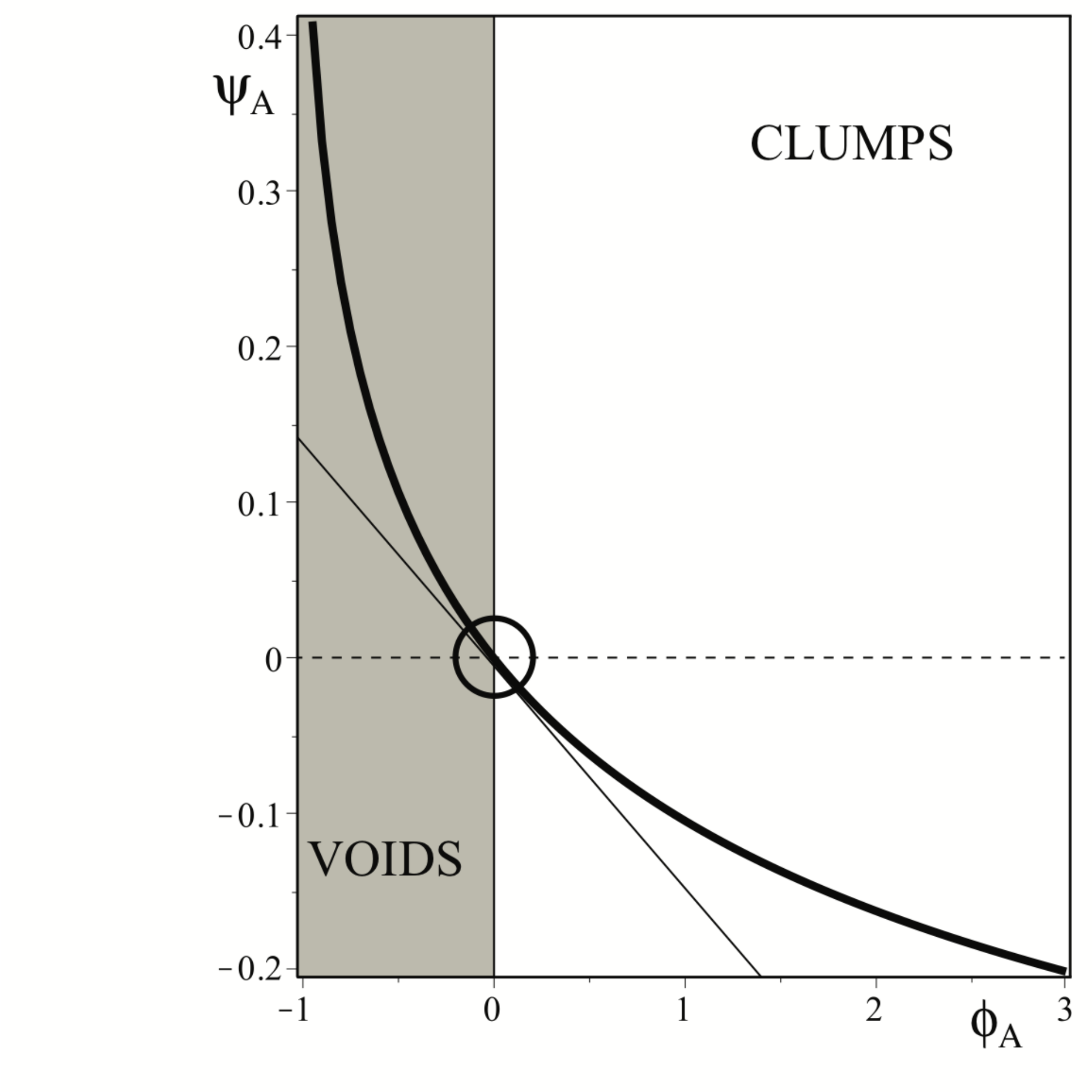}
\caption{{\bf Amplitudes of local and non--local perturbations.} The graph depicts the monotonic non--linear relation between the amplitude, $\phi_A$, of non--local perturbations and the maximal value of the local perturbation, $\psi_A$ (see figure 4), which shows that both perturbations provide a self--consistent qualitatively analogous description of the deviation from FLRW. Only for small amplitudes $|\phi_A|\ll 1$ the relation is linear.}
\label{fig5}
\end{center}
\end{figure}

\subsection{Amplitudes.}

The difference between the perturbations $\Da$ and $\Daav$ can also be understood in terms of the notion of ``amplitude'' often introduced in the context of simple intuitive Newtonian \cite{contrast1} and relativistic \cite{BoKrHe,kras2,BKHC2009} perturbations. In order to illustrate this point we consider an asymptotic perturbation in an LTB model converging radially to FLRW (as in figure 4), so that for all $A$ we have along the slices $\T[t]$ the asymptotically FLRW behavior of (\ref{asympt}):
\begin{equation} \mathop {\lim }\limits_{r_b \to \infty } A(r_b) = \mathop {\lim }\limits_{r_b \to \infty } \Aav_q[r_b] = A_{_\infty}(t)=\tilde A(t)\ne 0,\label{asflrw}\end{equation}
where $\tilde A(t)$ is the $A$ scalar for the asymptotic FLRW state (the exception is the spatial curvature if the FLRW state is Einstein de Sitter, as in this case $\KK_{_\infty}=\tilde\KK=0$). Assuming for simplicity a monotonic profile, then without loosing generality we can express any $A$ in any complete slice $\T[t]$ (not intersecting a singularity) in terms of a ``contrast amplitude'' $\phi_A$ as 
\ba \fl A = A_{_\infty}\left[1+\phi_{_A}\,f(r)\right],
\quad \hbox{with}\quad \phi_{_A}\equiv \Daav(0,\infty)=\mathop {\lim }\limits_{r_b \to \infty }\left[ \frac{A_c}{\Aav_q[r_b]}-1\right]=\frac{A_c}{A_{_\infty}}-1,\nonumber\\
\label{contrastq}\ea
where $A_c= A(t,0)=A_q(t,0)$ and $f$ is a smooth non--negative function satisfying $f_c=f(0)=1$ and $f'(0)=0$, as well as $f( r),\,f'( r)\to 0$ as $r\to\infty$.  Since we can always choose at any $\T[t]$ the radial coordinate as $R = r$, so that $R^2R'\dd r= r^2 \dd r$, the form (\ref{contrastq}) implies from I(13) and I(18)
\begin{equation}A_q( r)= A_{_\infty}\left[1+ \phi_A f_q( r)\right],\qquad f_q( r) =\langle f\rangle_q[r]=\frac{\int_0^r{f\,\bar r^2\dd \bar r}}{\int_0^r{\bar r^2\dd \bar r}},\end{equation}
while (\ref{Dadef}), (\ref{Danl}) and (\ref{deltas}) yield
\begin{equation} \Daav =\phi_{_A}\,f  \qquad
\Da = \frac{(f-f_q)\phi_{_A}}{1+\phi_A\,f_q},\label{EsDs}\end{equation}
which illustrate the following basic features: 
\begin{itemize}
\item The contrast amplitude $\phi_{_A}$ obeys a simple linear relation with the more intuitive  non--local perturbations $\Daav$, but its relation with local perturbations $\Da$ is non--linear. This is consistent with the fact that the latter perturbations are less intuitive than the former.   
\item Both perturbations vanish in the domain boundary $r\to\infty$ where the correspondence with the FLRW background occurs, but their behavior at the center is different: $\Daav(0,\infty)=\phi_{_A}\ne 0$ (in general) while $\Da(0)=0$ (see figures 4, 6, 8, 9 and 11).
\item 
The relation between the signs of the perturbations and the type of profile can also be given in terms of the amplitude: 
\bse\ba \fl \hbox{Clump profile:}\quad A_c>A_{_\infty},\quad \phi_{_A}>0,\qquad \Daav>0\quad \hbox{but}\quad \Da<0,\\
\fl \hbox{Void profile:}\quad\quad A_c<A_{_\infty},\quad \phi_{_A}<0,\qquad \Daav<0\quad \hbox{but}\quad \Da>0.\ea\ese
where we used the fact that (by construction) $f'\leq 0$ so that the sign relations (\ref{clump})--(\ref{void})  imply $f-f_q\leq 0$ and the sign of $\Da$ is the opposite of the sign of $\phi_A$.  
\end{itemize}
Evidently, the convention whereby a clump or void ({\it i.e.} overdensity or underdensity when $A=\rho$) respectively correspond to positive or negative contrast follows from the sign of the  amplitude contrast embodied in $\Daav$, and is more intuitive and easier to follow than the complicated relation between radial gradients $A'_q$ associated with $\Da$. 

However, in spite of their differences in signs when referred to clump/void radial profiles, the deviation form homogeneity (FLRW conditions) can be traced effectively with both types of perturbations, $\Daav$ and $\Da$,  because there is a consistent monotonic relation between their magnitudes (their maximal/minimal value in a given domain) and the amplitude $\phi_A$. For domains with monotonic $f$ the the amplitude $\phi_{_A}$ of the non--local perturbations $\Daav$ coincides with their extremal value, which occurs at the center $r=0$ (see figures 6b, 8b, 9b and 11b). For local perturbations the relation between $\phi_A$ and the extrema of $\Da$ (denoted by $\psi_{_A}$, see figure 4) is more complicated, as the latter may occur for different values of $r$ for different slices $\T[t]$ and depends on the choice of $f$ (see the maximum/minimum of the local perturbations in figures 6a, 8a, 9a and 11a). Since a result for a general $f$ is hard to find, we consider the special form $f=(1+r^n)^{-1}$ where $n>0$ (the results are qualitatively analogous for all other forms). The result, shown in figure 5, illustrates the non--linear monotonic relation between $\psi_{_A}$ and $\phi_A$. Hence, both types of perturbations provide a consistent estimate of the deviation from homogeneity of the models in which larger values of $|\phi_A|$ correspond to larger magnitudes of $\Da$ and $\Daav$.

\section{Linear limit.} 

While the $\Da$ and the $\Daav$ are exact quantities that need not be ``small' and do not comply with linear evolution equations, we can expect (intuitively) that they should somehow reduce to linear dust perturbations (in the comoving gauge) when their magnitudes ({\it i.e.} amplitudes) are small, as in the early times regime in the example of section 9. A more rigorous way to examine their connection to linear perturbations of dust sources follows by constructing second order equations for $\Drho$ and $\Drhoav$:
\begin{itemize}
\item {\bf Local perturbations}. We differentiate both sides of (\ref{EVq11}) and use the remaining equations (\ref{EVq12})--(\ref{EVq14}) to eliminate all derivatives except $\ddot\Drho$ and $\dot\Drho$, leading to:
\ba  \ddot\Drho -\frac{2\,[\dot\Drho]^2}{1+\Drho}+2\HH_q\,
\dot\Drho-4\pi \rho_q\,\Drho\left(1+\Drho\right)=0,\label{Dm2_eveq1}\ea
\item {\bf Non--local perturbations}. We follow the same procedure as above: differentiate (\ref{EVq31}) and use  (\ref{EVq32})--(\ref{EVq34}) to rewrite it in factors of $\dot\Daav$ and $\Daav$. Considering asymptotic backgrounds ($r_b\to\infty$) yields:
\ba  \fl\ddot\Drhoav -\frac{2\,[\dot\Drhoav]^2}{1+\Drhoav}+4\left(6\HH_{_\infty}-4\HH_q\right)\dot\Drhoav -\left[4\pi \rho_{_\infty}\Drhoav+6\left(\HH_q-\HH_{_\infty}\right)^2\right]\left(1+\Drhoav\right)=0,\nonumber\\\label{Dm2_eveq2}\ea
which holds for confined domains by replacing $\rho_{_\infty},\,\HH_{_\infty}$ with $\rhoav_q[r_b],\,\HHav_q[r_b]$. 
\end{itemize}
We remark that both (\ref{Dm2_eveq1}) and (\ref{Dm2_eveq2}) are exact non--linear equations for $\Drho$ or $\Drhoav$ (equations similar to (\ref{Dm2_eveq1}) have been obtained in \cite{kasai,ishak} for Szekeres models). Also, notice that (\ref{Dm2_eveq1}) and (\ref{Dm2_eveq2}) coincide as $r\to \infty$ for asymptotic perturbations and as $r\to r_b$ for confined domains $\DD[r_b]$.

In general there is no reason to assume that the amplitudes $\phi_A$ or their time derivatives $\dot\phi_A$ are small in a generic LTB model subjected to non--linear evolution equations, but if in a given range of $t$ we have $|\phi_A|\ll 1$ then the relation between both $\Da$ and $\Daav$ with $\phi_A$ should become similar. This can be seen by expanding for $|\phi_A|\ll 1$ in  the non--linear relation in  (\ref{EsDs}) between the $\Da$ and the amplitude $\phi_A$:  
\begin{equation} \Da \approx (f-f_q)\,\phi_A +O(\phi_A^2),\label{Dalin}\end{equation}
so that the $\Da$ become linear on $\phi_A$ (as the $\Daav$), and thus the magnitudes of $\Da$ and $\Daav$ are both proportional to $\phi_A$ because $0<f\leq 1$ and the fluctuations $f-f_q$ are bounded. As a consequence, if $|\phi_A|\ll 1$ both perturbations are of the same order in $\phi_{_A}$, and thus both comply with $|\Da|\ll 1$ and $|\Daav|\ll 1$ and  the time evolution of $\Da$ must be the same as that of $\Daav$ at leading order in $\phi_{_A}$. In particular, if we take $A=\rho,\,\HH$, the linearity conditions $|\phi_{_\rho}|\ll 1$ and $|\phi_{_\HH}|\ll 1$ imply that $|\Drho|\ll 1,\,\,|\Drhoav|\ll 1$ and $|\Dh|\ll 1,\,\,|\Dhav|\ll 1$ hold, and thus $|\dot\Drho|\ll 1,\, |\dot\Drhoav|\ll 1$ also holds (from (\ref{EVq13}) and (\ref{EVq31})). Therefore, we also have $\rho_q\approx\rho_{_\infty}\approx \rho$ and $\HH_q\approx\HH_{_\infty}\approx \HH$ (the same relations hold for finite domains by replacing $\rho_{_\infty},\,\HH_{_\infty}$ with $\rhoav_q[r_0],\,\HHav_q[r_0]$). Considering all these implications, if $|\phi_A|\ll 1$ the second order evolution equations (\ref{Dm2_eveq1}) and (\ref{Dm2_eveq2}) becomes at order $\phi_{_A}$
\ba  \ddot\Drho +2\HH_{_\infty}\,\dot\Drho-4\pi \rho_{_\infty}=0,\label{Dm2_eveqL}\ea
which is formally identical to the evolution equation for gauge invariant linear density perturbations of a dust source around a FLRW background characterized by $\rho_{_\infty},\,\HH_{_\infty}$ (or $\rhoav_q[r_b],\,\HHav_q[r_b]$ for bounded domains) in the comoving gauge~\cite{contrast1,contrast2}, which for dust is a synchronous gauge as well.

\section{A numerical example of a cosmic density void.}

In order to illustrate the utility of the evolution equations of section 4 for model building, as well as the properties and differences between local and non--local perturbations depicted qualitatively by figures 4b and 4d, we consider the evolution of a cosmic void configuration that emerges from the numerical solution of the systems (\ref{EVq11})--(\ref{EVq14}) and (\ref{EVq31})--(\ref{EVq34})
\footnote{We do not claim that this void model is ``realistic'' nor that it provides a good fitting to observations. Its purpose is simply to illustrate the evolution of the local and non--local perturbations of $\rho$ and $\HH$ and its relation to the radial profiles of these scalars.}
. For this purpose, we consider a regular hyperbolic model ($\KK_q< 0$ or $0<\Omega_q<1$) that is radially asymptotic to a spatially flat ($\tilde \KK=0$) FLRW model (Einstein de Sitter) \cite{RadAs} characterized by the Hubble factor, density and big bang time $\tilde H,\,\tilde\rho,\,\tilde\tbb$ that satisfy $8\pi\tilde\rho/(3\tilde H^2)=1$ and $\tilde H(t-\tilde\tbb)=2/3$ for all $t$. Hence, initial conditions must be selected such that shell crossings do not arise (see Appendix C of \cite{RadProfs}) and the limits $\KK_q\to 0$ or $\Omega_q\to 1$ as $r\to\infty$ hold along every time slice $\T[t]$ \cite{RadAs}.  

We consider the last scattering surface as the initial time slice, hence the subindex ${}_0$ will correspond to evaluation at $t=t_0=\tls$, which suggest using the constant $\tilde\Hls^{-1}=\tilde H^{-1}(\tls)$ as the characteristic length scale.  The nearly homogenous and spatially flat conditions at $\tls$ imply that initial value functions must satisfy:
\bse\ba  \HH_{q0}\approx \tilde\Hls,\quad \rho_{q0}\approx \tilde\rhols,\quad \tbb\approx \tilde\tbb,\quad \KK_{q0}\approx 0, \label{nearls1}\\
\Omega_{q0}=\frac{8\pi\rho_{q0}}{3\HH^2_{q0}}\approx 1,\quad \HH_{q0}(\tls-\tbb)\approx \frac{2}{3}, \label{nearls2}\ea\ese 
for all $r$, together with the strict limits $\rho_{q0}\to\tilde\rhols,\,\,\HH_{q0}\to \tilde\Hls,\,\,\Omega_{q0}\to 1$ and $\tbb\to \tilde\tbb$ as $r\to\infty$. By normalizing with respect to $\tilde\Hls$ we obtain the following dimensionless initial value functions that are compatible with (\ref{nearls1})--(\ref{nearls2}):
\bse\ba \fl \mu_{q0} \equiv \frac{4\pi\rho_{q0}}{3\tilde\Hls^2}=\frac{1}{2}\left[1+\frac{\epsilon}{1+ (r/r_1)^n}\right],\qquad \kappa_{q0}=\frac{\KK_{q0}}{\tilde\Hls^2}=-\frac{\zeta}{1+(r/r_2)^p},\label{initconds1}\\
\fl h_{q0}\equiv \frac{\HH_{q0}}{\tilde\Hls}=\left[2\mu_{q0}-\kappa_{q0}\right]^{1/2}=\left[1+\frac{\epsilon}{1+ (r/r_1)^n}+\frac{\zeta}{1+(r/r_2)^p}\right]^{1/2},\label{initconds2}\ea\ese 
where the positive constants $\epsilon,\,\zeta,\,n,\,p$ comply with $\epsilon\ll 1,\,\zeta\ll 1,\, n\leq 3,\,p\leq 2$ (the latter two follow from regularity conditions \cite{RadAs}), while $r_1,\,r_2$ are constant length scales, and we used the constraint (\ref{rHKq2}).  

Considering the initial value functions $\mu_{q0},\,h_{q0}$ in (\ref{initconds1})--(\ref{initconds2}) for the parameter values $n=3,\,p=3/2,\,r_1=1/2,\,r_2=1/\sqrt{3},\,\epsilon =0.001,\,\zeta=0.01$, we integrate the system (\ref{EVq11})--(\ref{EVq14}) for the dimensionless variables $\mu_q=4\pi\rho_q/(3\tilde\Hls^2),\,h_q=\HH_q/\tilde\Hls,\,\Drho,\,\Dh$ in terms of the dimensionless time
\begin{equation} \tau \equiv \tilde\Hls(t-\tbb)-\frac{2}{3},\label{taudef}\end{equation} 
so that $\tau=0$ corresponds to $t=\tls$, since $\tbb\approx \tilde\tbb$ with $\tilde\tbb-\tbb(0)\sim O(\hbox{max}(\epsilon,\zeta))$. The bang time is marked by $\tau\approx -2/3+O(\hbox{max}(\epsilon,\zeta))$, though the LTB model is no longer valid for $t<\tls$ (or $-2/3<\tau<0$), and thus we only consider the range $\tau\geq 0$. Since $\tilde\Hls=\tilde H_0(1+z)^{3/2}$ and $z(\tls)\sim 1100$, the present value of $\tilde H_0\approx 70\,\hbox{km}/(\hbox{sec Mpc})$ yields for the present time the value $\tau =3.6\times 10^4$, which approximately corresponds to 13 Gys. 

By considering non--local perturbations that are asymptotic ({\it i.e.} $r_b\to\infty$ taking the form (\ref{asdeltaNL})), together with a FLRW background state that is Einstein de Sitter, the system (\ref{EVq31})--(\ref{EVq34}) simplifies considerably, since the terms $\rhoav_q[r_b],\,\HHav_q[r_b]$ in these equations take the asymptotic forms $\tilde\rho,\,\tilde H$ that comply with $\Omav_q=\tilde\Omega=1$. Hence, all terms involving $\rhoav_q[r_b],\,\HHav_q[r_b]$ can be replaced by closed exact analytic dimensionless forms $\tilde \mu,\,\tilde h$ associated with the Einstein de Sitter asymptotic state, which can be given analytically in terms of the dimensionless time (\ref{taudef}) by: 
\begin{equation} \fl \tilde h = \frac{\tilde H}{\tilde \Hls}=\frac{1}{\tilde a^{3/2}},\qquad  \tilde \mu=\frac{4\pi\tilde\rho}{3\Hls^2}\frac{\tilde h}{2}=\frac{1}{2\tilde a^3},\qquad \tilde a=\left[1+\frac{3}{2}\tau\right]^{2/3}.\label{mh}\end{equation}
However, from a computational point of view, it is easier to obtain the non--local perturbations  through the relation (\ref{deltas}):
\begin{equation}\Drhoav = \frac{\mu_q}{\tilde \mu}\left(1+\Drho\right)-1,\qquad \Dhav = \frac{h_q}{\tilde h}\left(1+\Dh\right)-1,\end{equation}
where $\tilde \mu,\,\tilde h$ are given by (\ref{mh}) and $\mu_q,\,h_q,\,\Drho,\,\Dh$ follow from the numerical integration of (\ref{EVq11})--(\ref{EVq14}).
\begin{figure}
\begin{center}
\includegraphics[scale=0.35]{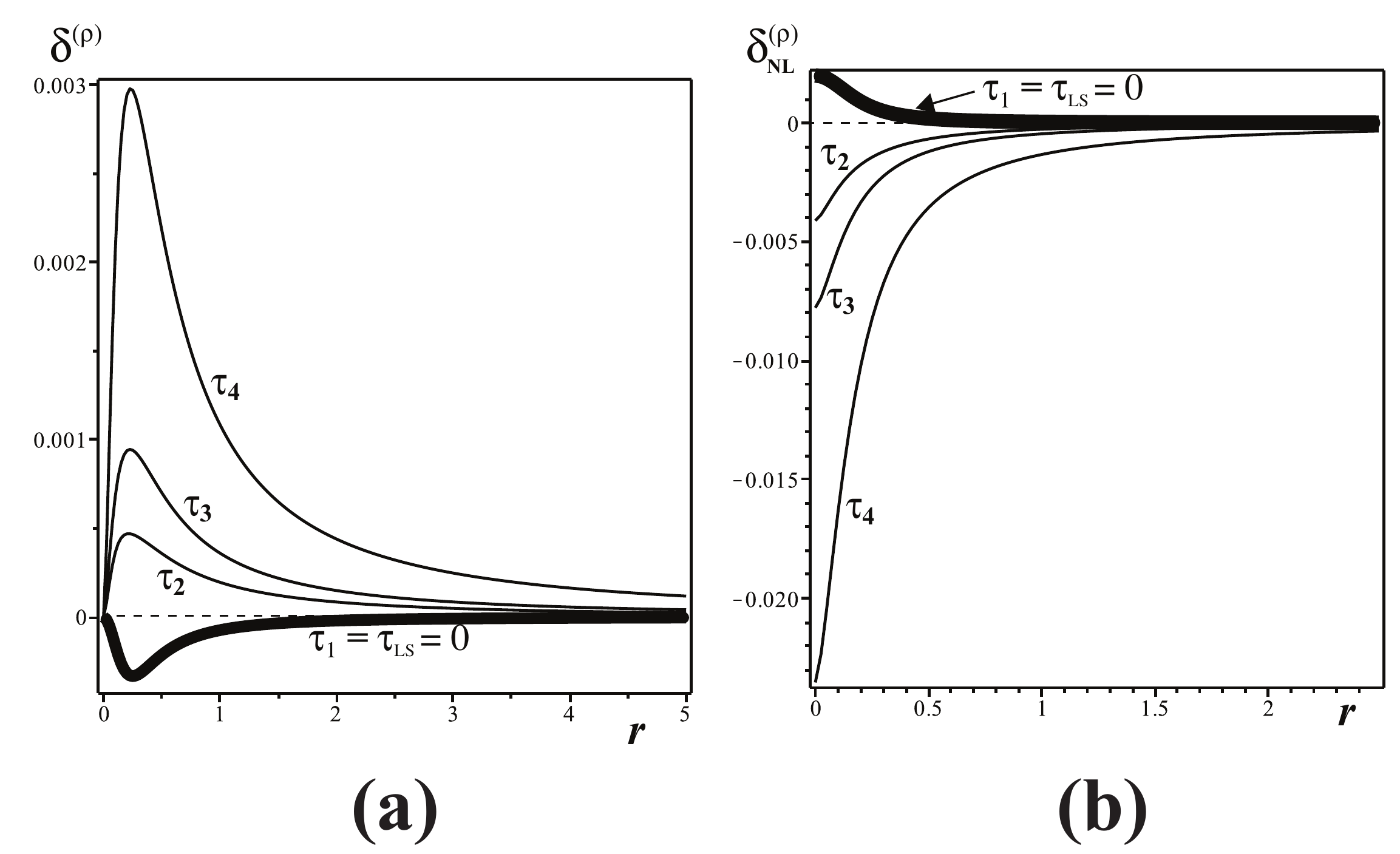}
\caption{{\bf Radial profile of density perturbations (early times).} Panels (a) and (b) respectively depict local and non--local perturbations for values of $\tau$ corresponding to early cosmic times $\tau_i,\,i=1,2,3,4$ corresponding to $0,0.5,1.0,5.0$. The profiles have the same form as depicted in the qualitative diagrams of figures 4b and 4d. The thick curves mark the last scattering surface $\tau=\tau_1=0$. Notice how the perturbations invert their sign between $\tau_1$ and $\tau_2$, corresponding to the density profile inversion shown in figure 7a (see the text for further explanation).}
\label{fig6}
\end{center}
\end{figure}
\begin{figure}
\begin{center}
\includegraphics[scale=0.35]{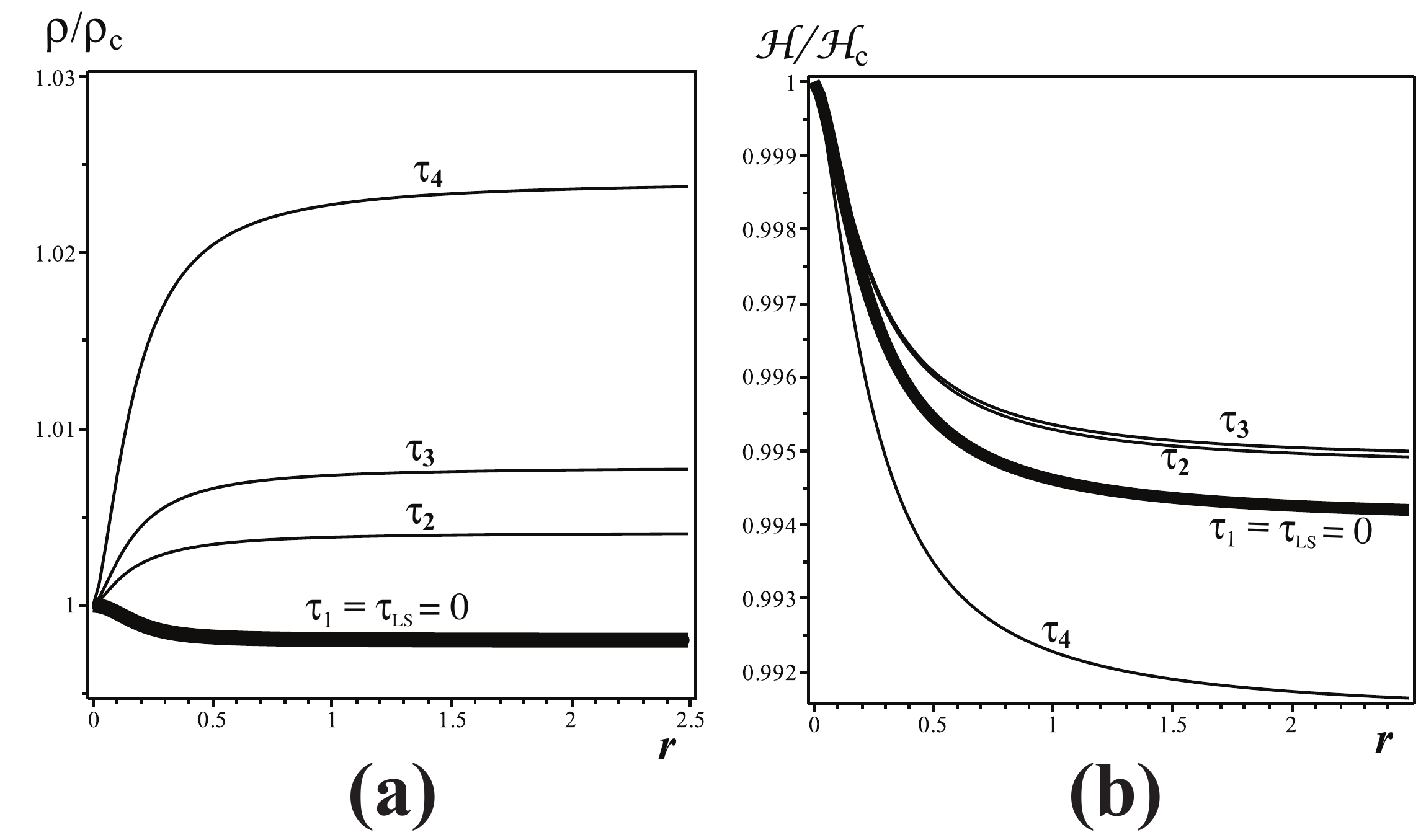}
\caption{{\bf Radial profiles of the density and Hubble scalar (early times).} Radial profiles of the local scalars $\rho$ and $\HH$ normalized by their central values $\rho_c=\rho(t,0)$ and $\HH_c=\HH(t,0)$ for the same values of $\tau$ of figure 6. The profiles have the form sketched in figures 4a and 4c.  Notice the transition from a clump to a void profile of $\rho$ between $\tau_1$ and $\tau_2$, while $\HH$ has a clump profile for all $\tau$. }
\label{fig7}
\end{center}
\end{figure}
\begin{figure}
\begin{center}
\includegraphics[scale=0.35]{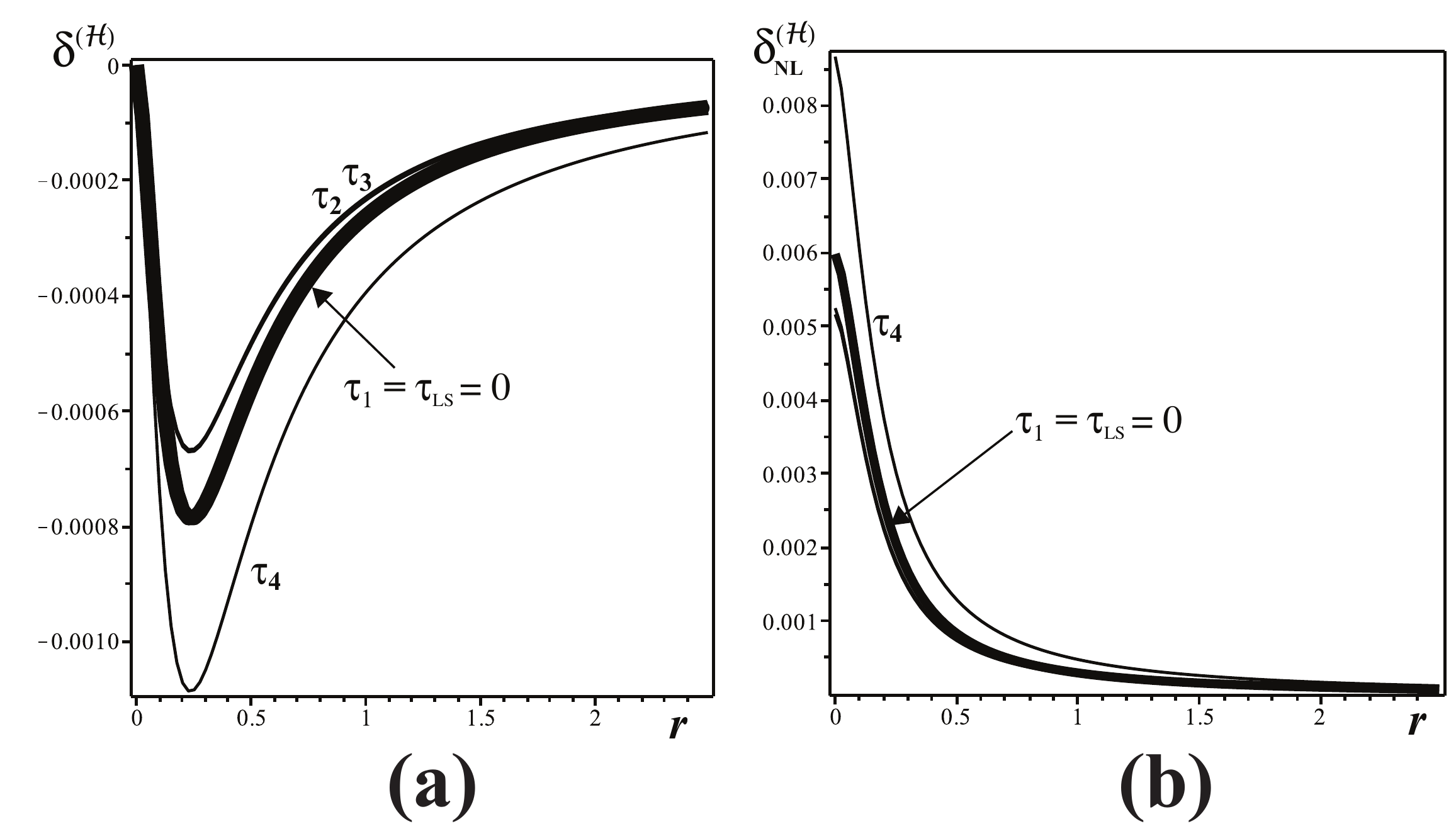}
\caption{{\bf Radial profiles of perturbations of the Hubble scalar (early times).} Panels (a) and (b) respectively depict local and non--local perturbations for the same values of $\tau$ of figure 6. The sign of the latter is characteristic of a clump profile (see figure 7b).}
\label{fig8}
\end{center}
\end{figure}
\subsection{Early cosmic times.}

Since $\Delta\tau= 1$ corresponds to a time lapse $\tilde\Hls^{-1} \sim 10^5 \hbox{ys}$, values $\tau\sim O(1)$ should describe a linear regime with small amplitudes for both local and non--local perturbations. This can be appreciated in figures 6 and 8, depicting the radial profiles of local and non--local perturbations of the density and the Hubble scalar for the values $[\tau_1,\,\tau_2,\,\tau_3,\,\tau_4]=[0,\,0.5,\,1.0,\,5.0]$ that range between the last scattering surface ($\tau=0$, marked by thick curves) to about $10^6\,\hbox{ys}$ \,($\tau=5$). 

The difference between local and non--local perturbations and its relation to the perturbation amplitude and type of radial profile (clump or void) clearly emerges by comparing density perturbations in figures 6a and 6b and the radial density profiles in figure 7a. Notice how the amplitudes of the perturbations, $\phi_\rho$ and $\psi_\rho$, exhibit the behavior described in section 7 and depicted in figure 4. As shown in figure 7a, the initial configuration is a clump with small amplitude (thick curve marked by $\tau=0$): it is a positive non--local initial perturbation  ({\it i.e.} positive density contrast) depicted by the thick curve in figure 6b, while the initial local perturbation (thick curve in figure 6a) is negative (negative radial gradient of the density). As the evolution proceeds, both types of density perturbation in figures 6a and 6b have reversed their sign at $\tau=\tau_2$, with the profile passing from that of a clump to a void in figure 7a
\footnote{The occurrence of this profile inversion in regular hyperbolic models is reported in \cite{RadProfs}. A necessary condition for it is $\Drho_{q0}-(3/2)\DKK_{q0}\geq 0$.}
:  the non--local perturbation becomes negative (voids have negative density contrast) and the local one becomes positive (radial gradient of density is positive in voids). However, the perturbations of the Hubble scalar in figures 8a and 8b do not change sign: the non--local perturbation is positive (positive contrast) and the local one is negative (negative radial gradient), and thus the radial profile is that of a clump for all these values of $\tau$ (see figure 7b).
\begin{figure}
\begin{center}
\includegraphics[scale=0.35]{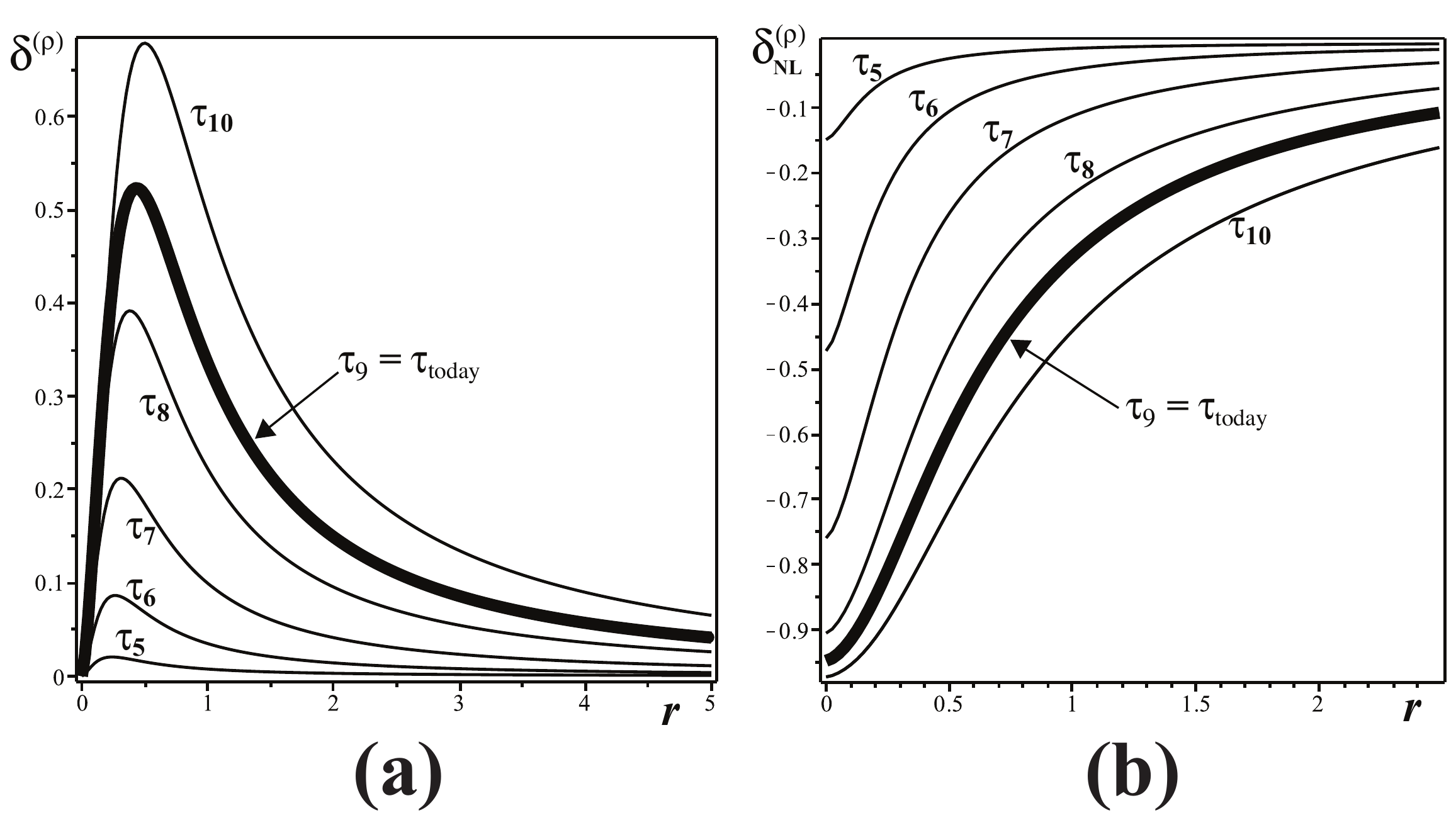}
\caption{{\bf Radial profile of density perturbations (late times).} The panels respectively depict local and non--local perturbations for time values $\tau_i,\,i=1,2,3,4,5,6,7,8,9,10$ given by $100,\,1000,\,5000,\,18000,\,36000,\,72000$, with the thick curve marking present day cosmic time $\tau_9=36000$, which  roughly corresponds to $13$ Gys. Local perturbtions have positive sign while the non--local ones are negative, all of which is characteristic of void profiles (see figure 10a). The amplitudes clearly correspond to the non--linear regime. }
\label{fig9}
\end{center}
\end{figure}
\begin{figure}
\begin{center}
\includegraphics[scale=0.35]{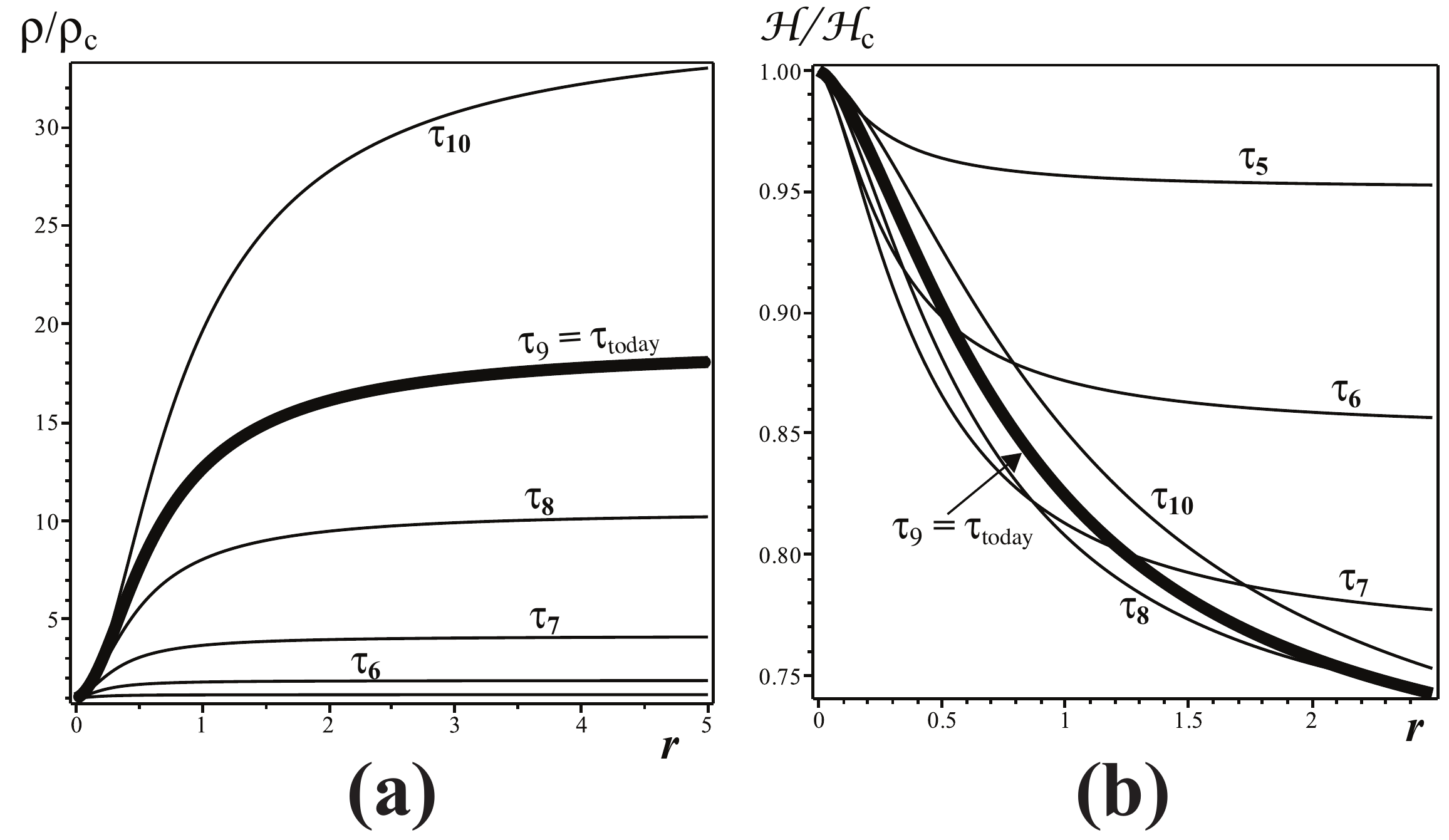}
\caption{{\bf Radial profiles of the density and Hubble scalar (late times).} The radial profiles of the same scalars of figure 7 for same cosmic times as figure 9. Notice the density void profile, while the Hubble scalar keeps a clump profile.}
\label{fig10}
\end{center}
\end{figure}
\begin{figure}
\begin{center}
\includegraphics[scale=0.35]{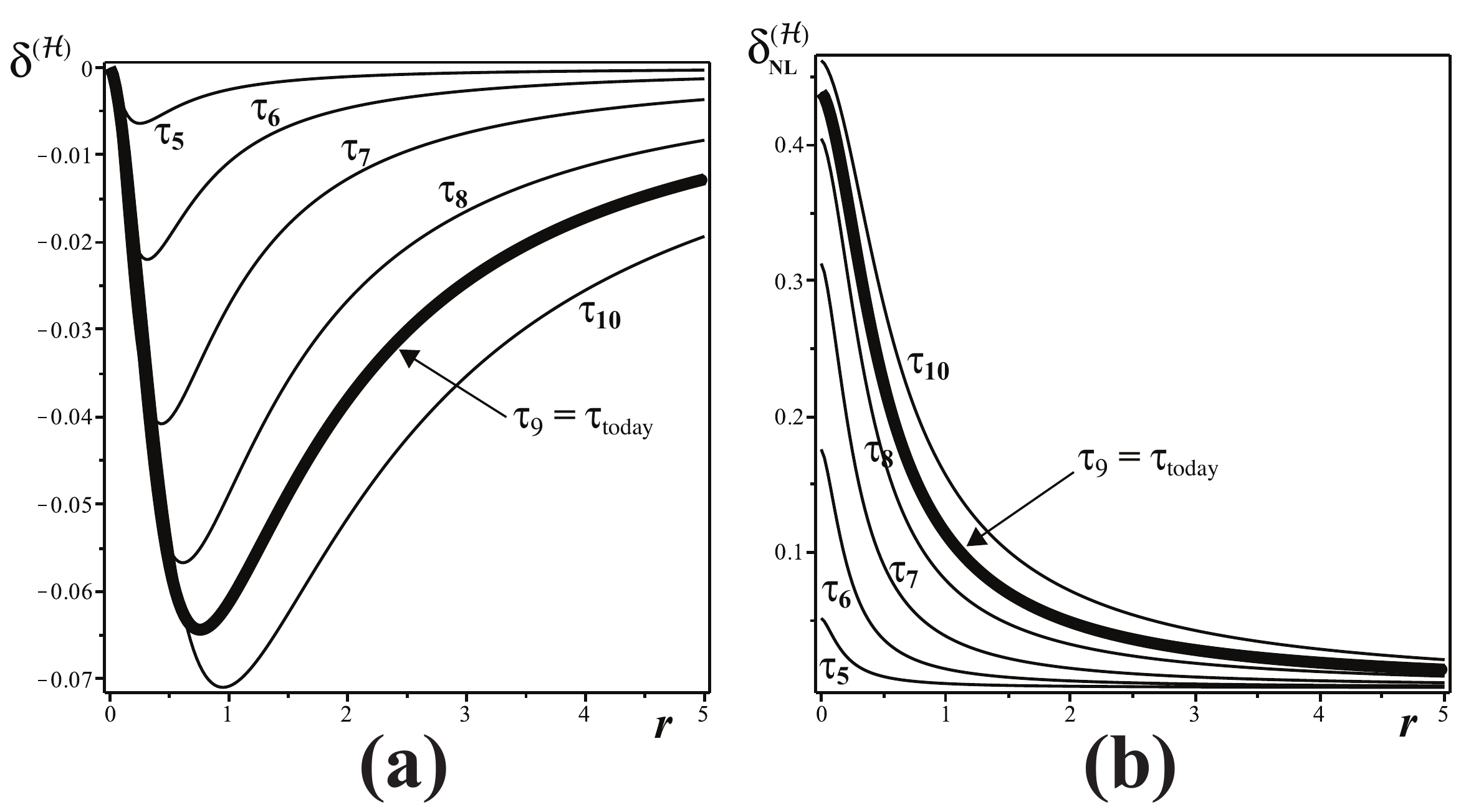}
\caption{{\bf Radial profiles of perturbations of the Hubble scalar (late times).} The same perturbations of figure 8 for late cosmic times. The signs of the perturbations are characteristic of a clump profile. While the amplitude of the local perturbation remains small, the amplitude of the non--local perturbation is large. }
\label{fig11}
\end{center}
\end{figure}
\subsection{Late cosmic times.} 

We consider now cosmic times $\tau_5,\tau_6,\tau_7,\tau_8,\tau_9,\tau_{10}$ respectively corresponding to $[100,\,1000,\,5000,\,18000,\,36000,\,72000]$. Hence the curve marked by $\tau_9=36000$ approximately corresponds to present day cosmic time and is depicted by thick curves. The radial profiles of local and non--local density perturbations are displayed for these values of $\tau$ by figures 9a and 9b. Evidently, the perturbations' amplitude increases to large values that  indicate a non--linear regime and the density profile is that of a density void (figure 10a). The sign of the non--local perturbations is negative, as the density contrast is negative in voids, reaching an over $\sim 90$\% negative contrast at the center for late cosmic times (as in the void models of chapter 4 of \cite{BKHC2009}). However,  local perturbations are positive, as radial gradients are positive in void profiles. On the other hand, as shown by figures 11a and 11b, the local and non--local perturbations of the Hubble scalar are, respectively, negative and positive, characteristic of a clump profile (figure 10b). While the amplitudes of the local perturbations of the Hubble scalar remain small and within linear regime ($\sim 10^{-2}$), the amplitudes of these perturbations show a steep growth from figure 9a to 11a. On the other hand, the amplitudes of the non--local perturbations of the Hubble scalar become large, showing a present day $\sim 30$ \% positive contrast. 
\begin{figure}
\begin{center}
\includegraphics[scale=0.35]{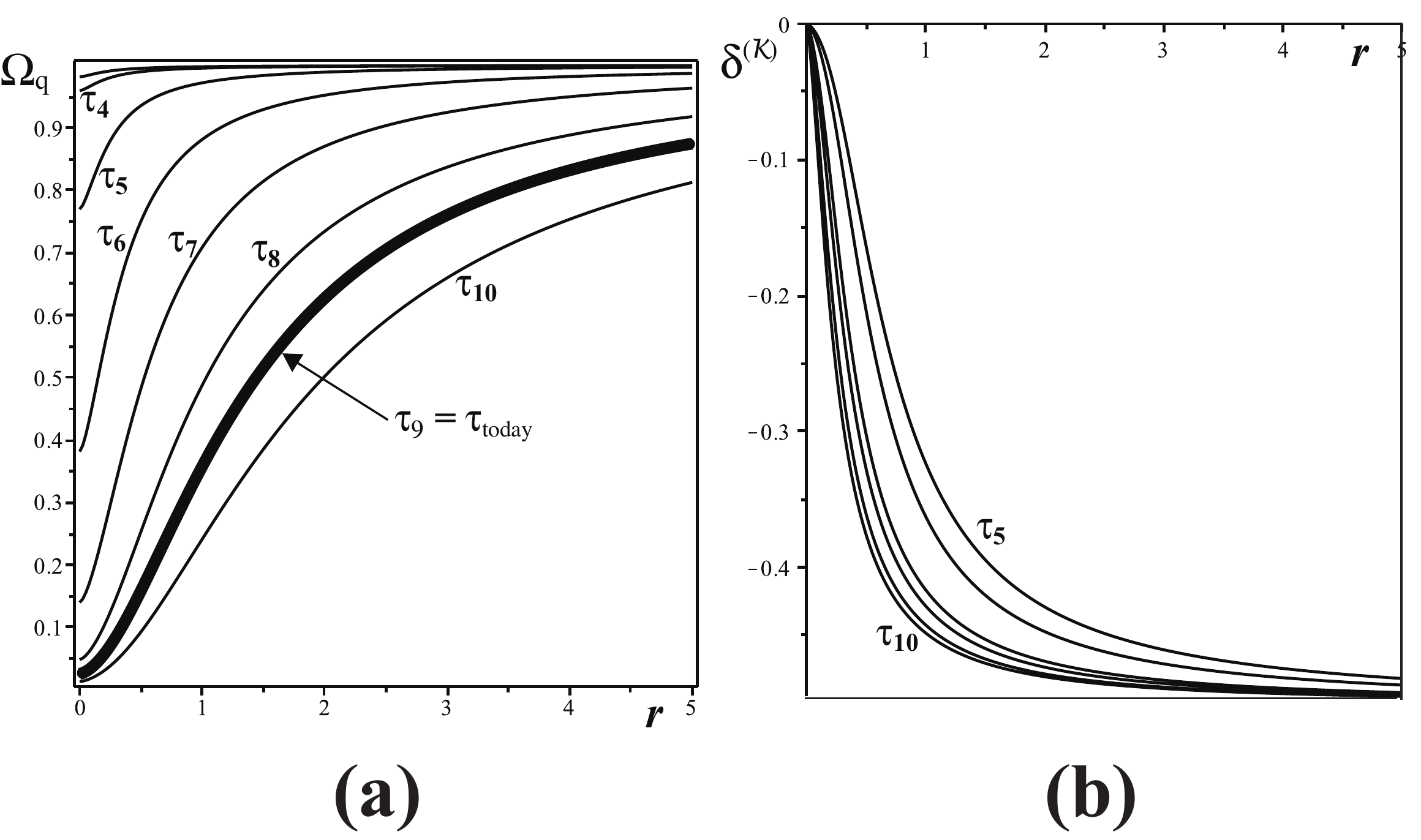}
\caption{{\bf Radial profile of $\Omega_q$ and spatial curvature perturbation $\DKK$.} Panel (a) displays the radial profile of the q--scalar $\Omega_q$, defined by (\ref{Omdef}). The curves shows the transition from nearly homogeneous spatially flat conditions at $t=\tls$ ($\Omega_q\approx 1$) to a very inhomogeneous void with negative curvature with $\Omega_q\ll 1$ at the void center in the present day cosmic time. Panel (b) depicts the local perturbation of the spatial curvature $\DKK$. Since $\KK\to 0$ as $r\to\infty$, this perturbation is large with $\DKK\to -1/2$ in this limit.}
\label{fig12}
\end{center}
\end{figure}

The radial profiles of the q--scalar $\Omega_q$, depicted in figure 12a, describe the evolution of this LTB model as a transition from nearly homogeneous and spatially flat conditions ($\Omega_q\approx 1$) at $t=\tls$ to a very inhomogeneous void with $\Omega_q\ll 1$ with marked negative spatial curvature. Since the negative spatial curvature complies with $\KK_q\to 0$ for all $t$, the non--local spatial curvature perturbations cannot be defined because $\KKav_q\to\tilde\KK=0$, but there is no problem in defining local perturbations $\DKK$. As shown in \cite{RadAs}, the local perturbation of a scalar tending asymptotically to zero is not small. For the initial conditions (\ref{initconds1})--(\ref{initconds2}), $\DKK$ reaches an asymptotic limiting value $\DKK\to -1/2$ as $r\to\infty$ (see figure 12b).       

\section{Conclusion and summary.}

We have examined the perturbations that emerge from the weighed averaging formalism developed in part I. These perturbations can be local ($\Da$) or non--local ($\Daav$), depending on whether they are, respectively, defined for q--scalars that are functions or functionals (a summary of part I is given in section 2). By introducing an initial value parametrization, we showed in section 3 that the q--scalars common with FLRW models ($A_q$ and $\Aav_q[r_b]$ for $A=\rho,\,\HH,\,\KK,\,\Omega$) identically satisfy FLRW scaling laws that mimic FLRW expressions commonly used in LTB void models that probe the possibility of explaining observations without resorting to dark energy \cite{obs1,obs2,kolb,GBH,alnes,brouzakis,bismanot,swisscheese,endqvist,bisnotval,clarkson,marranot}. Also, the q--scalars determine (via Darmois matching conditions) a unique FLRW background state that can be either a formal reference FLRW model that is different for each domain, or an actual background FLRW spacetime if a domain $\DD[r_b]$ is smoothly matched to a FLRW region (Swiss cheese models).  The following results are worth highlighting:
\begin{itemize}
\item As proven in part I (see I(42a)--I(42b) and I(43)), the local density perturbation $\Drho$ is directly expressible in terms of the ratio of Weyl to Ricci curvature: $\Psi_2/\CR$, where $\Psi_2$ is the only nonzero Newman--Penrose conformal invariant and $\CR$ is the Ricci scalar, whereas $\Dh$ is expressible in terms of the ratio of anisotropic vs. isotropic expansion: $\Sigma/\HH$, where $\Sigma$ is the eigenvalue of the shear tensor. All q--scalars can be given in terms of these scalar invariants by means of the constraints (\ref{Dhdef})--(\ref{DOmdef}). 
\item The evolution equations for the q--scalars and their perturbations (local and non--local)  completely determine the dynamics of the models, and thus provide an alternative to the use of analytic solutions in the study and applications of the models. Although the resulting systems involve PDE's, they contain only time derivatives and the constraints are purely algebraic in nature and preserved in time. Hence, these equations can be treated effectively as ODE's (section 4). We tested the numerical integration of these systems in the example provided in section 9 of a void model that is radially asymptotic to an Einstein de Sitter FLRW model.   
\item Since the q--scalars behave effectively as FLRW scalars and their perturbations convey for every domain  the (exact) deviation from FLRW dynamics, we can  rigorously re--interpret LTB dynamics as the dynamics of exact spherical dust perturbations  on a FLRW background defined by the q--scalars. As mentioned before, this background state is an abstract reference background defined at every domain by the continuity of the q--scalars under Darmois matching conditions (\ref{mcon1})--(\ref{mcon3}). It is an actual FLRW spacetime only if a smooth matching with a FLRW region is considered in the context of ``Swiss cheese'' models  (section 5).
\item Both local and non--local fluctuations and perturbations can be either confined in a given domain $\DD[r_b]$ (with or without assuming a ``Swiss cheese'' configuration through a matching with a FLRW region, see figures 1, 2 and 3), or an asymptotic perturbation for the case when $\DD[r_b]$ becomes the whole time slice $\T[t]$ in the limit $r_b\to\infty$ (see figure 4). The choice between confined and asymptotic perturbations depends on the boundary conditions of the specific problem or application that we may work out with LTB models:
\begin{itemize}
\item Confined perturbations (local or non--local) without a matching with FLRW are practical because they are easily applicable to study the inhomogeneity pattern of any generic LTB configuration, but are more abstract because the FLRW background state is a fictitious reference spacetime that changes for each domain. 
\item Swiss cheese configurations are more intuitive because the background state is a single (and non--fictitious) FLRW spacetime. They are useful if we wish to describe various inhomogeneous regions, as in Swiss cheese void models used to fit observations \cite{kolb,brouzakis,bismanot,swisscheese} (see reviews in \cite{BKHC2009} and \cite{marranot}), but their intuitiveness is offset by the emergence of artificial jointly packed layers near the hole boundary (``humps'' and ``bags'' in the radial profile (see figures 1, 2 and 3) that arise because of need to impose continuity of the perturbations, which must vanish in a non--fictitiuos background located at a finite radius (also, shell crossings necessarily arise if the LTB region is hyperbolic, see Appendix A).
\item Asymptotic perturbations do not present these inconveniences, and thus are more natural, and also practical if the radial profile of the scalars rapidly converges to a given set of FLRW values (as is the case in the model presented in section 9 and illustrated by figures 6--12).
\end{itemize}
\item The local perturbations provide a measure of inhomogeneity through the local magnitude of the radial gradients of covariant scalars, while non--local perturbations do so through the familiar notion of the ``contrast'' between local values of these scalars and a reference FLRW value defined as a q--average for each fixed whole domain, or for the whole slice in the asymptotic limit. Even if their evolution equations are more complicated, non--local perturbations are more intuitive than local ones, as they allow us to associate over and under densities with (respectively) positive or negative density amplitudes, while local perturbations yield the opposite sign. We have illustrated these points through the notion of the ``amplitude'' of the perturbations (see sections 6 and 7 and figure 5 and compare the profiles of $\rho,\,\HH$ in figures 7 and 10 with the profiles of their local perturbations in figures 6a, 8a, 9a and 11a). As a possible application, the exact local and non--local perturbations  may be useful in understanding structure formation through an approach that involves the radial gradient and profile of the perturbations (as for example in \cite{hidalgo}).   
\item The definition of non--local perturbations $\Daav$ by the maps (\ref{map1a}) and (\ref{map2a}) provide (through the q--average and its relation to kinematic and curvature invariants) a rigorous and coordinate independent interpretation for simple examples of perturbations that are based on the notion of a contrast with respect to a FLRW background. These simple contrast perturbations are frequently introduced as ansatzes in many text--book and articles, for example, in the context of simple Newtonian models of structure formation (the ``top hat'' or  ``spherical collapse model'' \cite{contrast1}) but also in linear perturbations \cite{contrast2} and in astrophysical applications of LTB models (see examples in \cite{BoKrHe,kras2,BKHC2009}).  
\item Both local and non--local perturbations yield in the linear limit the familiar dust perturbations of linear theory in the comoving (or isochronous) gauge (section 8).
\item The example of a void model given in section 9 (which follows from the numerical integration of the evolution equations of section 4) clearly illustrate the properties of local and non--local perturbations described above. It also shows the potential utility of the q--scalars and their perturbations as tools for LTB model building.       
\end{itemize}     

While the q--average formalism cannot be used to study and understand the dynamical relation between perturbations and back--reaction, it does provide through the results of part I and the present article (part II) interesting theoretical connections between averaging, perturbation theory, invariant scalars and statistical correlations of $\rho$ and $\HH$, which signals a valuable theoretical insight on how the averaging process should work in any generic solution of Einstein's equations (at least in LRS spacetimes whose dynamics is reducible to scalar modes). A coordinate independent study and generalization of the ``growing'' and ``decaying'' models \cite{wainwright} is still necessary for a complete theoretical study of the perturbations that we have examined. This task will be undertaken in a separate article currently under preparation.  

Since most formal and theoretical results obtained for LTB models can be readily applied to Szekeres models \cite{sussbol}, the same type of exact perturbation formalism can be devised for these non--spherical models and perhaps even to more general spacetimes. Specifically, we propose identifying fluid flow  scalars that comply with FLRW dynamics (hopefully related to some weighed average), and then expressing local 1+3 scalars as fluctuations of the new scalars. The following step would be to explore if the 1+3 evolution equations given in terms of the new set of variables has the structure of evolution equations and constraints in the context of a formalism of exact perturbations on a FLRW background. We feel that this proposal is worth considering in future research. 

\section*{Acknowledgements.} 

The author acknowledges financial support from grant SEP--CONACYT 132132. Acknowledgement is also due to Thomas Buchert for useful comments.           

\begin{appendix}

\section{Shell crossings and Swiss cheese models.} 

In a Swiss cheese model in which the FLRW background is a spatially flat Einstein de Sitter model the second of the Darmois matching conditions (\ref{mcon1})--(\ref{mcon2}) and the supplementary condition (\ref{mcon4}) must be replaced by:
\begin{equation}\fl \KK_{qb}=\KKav_{q}[r_b]=\tilde\KK(t)=0,\quad \Omega_{qb}=\Omav_{q}[r_b]=\tilde\Omega(t)=1,\quad \DKK_b =0.\label{Ap1}\end{equation}
where $r_b$ is bounded. We prove in this Appendix that these conditions are incompatible with the following condition to prevent shell crossings in hyperbolic models $\KK_{q0}<0$ (see \cite{ltbstuff} and Appendix C of \cite{RadProfs}): 
\begin{equation}\DKK_0 = \frac{rE'}{3E}-\frac{2}{3}\geq -\frac{2}{3}\quad \Leftrightarrow\quad E'=(|\KK_{q0}|r^2)'\geq 0.\label{Ap2}\end{equation}
In order to comply with (\ref{Ap1}) we must have  $|\KK_{q0}|\to 0$ as $r\to r_b$, and thus $|\KK_{q0}|r^2\to 0$ must also hold in this limit, but (\ref{Ap2}) requires $|\KK_{q0}|r^2$ to be monotonically increasing in the range $0\leq r\leq r_b$ and we have $E=|\KK_{q0}|r^2 \to 0$ as $r\to 0$ \cite{ltbstuff,kras2,BKHC2009,RadAs}. Therefore, $E'$ necessarily reverses its sign in this range. As a consequence, shell crossings necessarily occur in all Swiss cheese models with an Einstein de Sitter background in which the LTB section is hyperbolic (regardless of the type of profile of the density). This problem, which has been reported by \cite{kolb} and commented in chapter 5.3.5 of \cite{BKHC2009}, does not arise if the LTB region is elliptic (as the sign of $E'$ is not constrained) and in hyperbolic models radially converging asymptotically to an Einstein de Sitter state (as the void model of section 9), since in this case functions $\KK_{q0}$ can be found such that both $|\KK_{q0}|\to 0$ as $r\to \infty$ and $(|\KK_{q0}|r^2)'\geq 0$ hold \cite{RadAs}: for example, if $|\KK_{q0}|$ decays as $r^{-p}$ with $0<p<2$ (as we assumed in (\ref{initconds1})).     

\end{appendix}

\end{document}